\documentclass{article}

\usepackage{graphicx}
\usepackage{fullpage,amsmath,amssymb,latexsym}
\usepackage{multirow}
\usepackage{natbib}

\begin{document}

\title{Shapes and gravitational fields of rotating two-layer Maclaurin ellipsoids: Application to planets and satellites}

\author{Gerald Schubert $^{1}$, John D. Anderson$^{2}$, Keke Zhang$^3$, Dali Kong$^{3}$, and Ravit Helled$^1$\\
\small{$^1$Department of Earth and Space Sciences}\\
\small{University of California, Los Angeles, CA 90095Ð1567, USA}\\
\small{$^2$Jet Propulsion Laboratory \footnote{Retiree},} \\
\small{4800 Oak Grove Drive, Pasadena, CA 91109}\\
\small{$^3$ Department of Mathematical Sciences} \\
 \small{College of Engineering, Mathematics and Physical Sciences}\\
  \small{University of Exeter, Exeter, UK}\\
}

\date{}
\maketitle

\begin{abstract}
The exact solution for the shape and gravitational field of a rotating two-layer Maclaurin ellipsoid of revolution
is compared with predictions of the theory of figures up to third order in the small rotational parameter of the theory
of figures.
An explicit formula is derived for the external gravitational coefficient $J_2$  of the exact solution.
A new approach to the evaluation of the theory of figures based on numerical integration of ordinary differential equations is presented.
The classical Radau-Darwin formula is found not to be valid for the rotational parameter $\epsilon_2 = \Omega^2/(2\pi \mathrm{G}\rho_2) \geq 0.17$
since the formula then predicts a surface eccentricity that is smaller than the eccentricity of the core-envelope boundary.
Interface eccentricity must be smaller than surface eccentricity. In the formula for $\epsilon_2$, $\Omega$
is the angular velocity of the two-layer body, $\rho_2$ is the density of the outer layer,
and G is the gravitational constant.  For an envelope density of 3000~kg~m$^{-3}$ the failure of the Radau-Darwin
formula corresponds to a rotation period of about 3~hr. Application of the exact solution and the theory of figures is made to models of Earth,
Mars, Uranus, and Neptune. The two-layer model with constant densities in the layers can provide realistic approximations to
terrestrial planets and icy outer planet satellites. The two-layer model needs to be generalized to allow for a continuous
envelope (outer layer) radial density profile in order to realistically model a gas or ice giant planet.
\end{abstract}


\section{Introduction}
\label{sec:intro}
\citet{KongZhangSchub:JGR2010}
have presented an exact theory for the rotational distortion of a rotating two-layer spherical body with a constant density core
surrounded by an envelope (outer layer) with a different constant density.
The solution for the case when the core and envelope have equal densities, the
Maclaurin ellipsoids, was obtained more than 250 years ago and attracted the attention of such notables as d'Alembert,  Clairaut, Euler, Laplace, Legendre,
Poisson and Gauss. The solutions are discussed in
\citeauthor{Chand:EFE1969} (\citeyear{Chand:EFE1969}, Ellipsoidal Figures of Equilibrium)
and
\citeauthor{Lamb:H1932}~(\citeyear{Lamb:H1932}, Hydrodynamics).
Until the work of
\citet{KongZhangSchub:JGR2010},
the classical solution for the constant density Maclaurin ellipsoid had not been generalized to a body with
non-uniform density. Instead, approximate solutions, for bodies with density increasing from the surface
to the center, have been developed by geophysicists interested in the internal structures of the Earth and
planets. The approximate solutions fall under the umbrella of the theory of figures
\citep[][Physics of Planetary Interiors]{ZharkTrubi:PPI1978}
and rely on the smallness of a rotational parameter that measures the distortion of a rotating body. In this
paper we refer to the theory developed by
\citet{ZharkTrubi:PPI1978}
for the shapes of fluid bodies as the theory of figures.
The rotational distortions of the Earth and planets are indeed small, but it is important to accurately determine
the small distortions to correctly infer the interior structures of the bodies. Accordingly, the theory of figures has been
developed to high order in the small rotational parameter.\par

While the two-layer spherical body with constant core and envelope densities is too simple a model for the careful study of most planets,
the exact solution for the rotational distortion of such a body provides a benchmark against which the accuracy and range of validity
of approximate solutions and numerical models can be evaluated. In this paper we compare results from the exact solution
for the rotating two-layer Maclaurin ellipsoids to those obtained using the Radau-Darwin approximation and other simplifications
based on the theory of figures. We derive formulae for the rotationally distorted two-layer sphere using the theory of figures valid to
third order in the small rotational parameter and assess these results against the exact solution. Application of the exact theory
is made to the Earth and planets while keeping in mind the limitations of using a two-layer model to represent these bodies.
Both the theory of figures and the exact theory of
\citet{KongZhangSchub:JGR2010}
assume hydrostatic equilibrium of the rotating body.\par

\section{Exact Solution for a Rotating Two-Layer Spherical Body}
\label{sec:twolayersol}
\begin{figure}
\includegraphics[scale=.75]{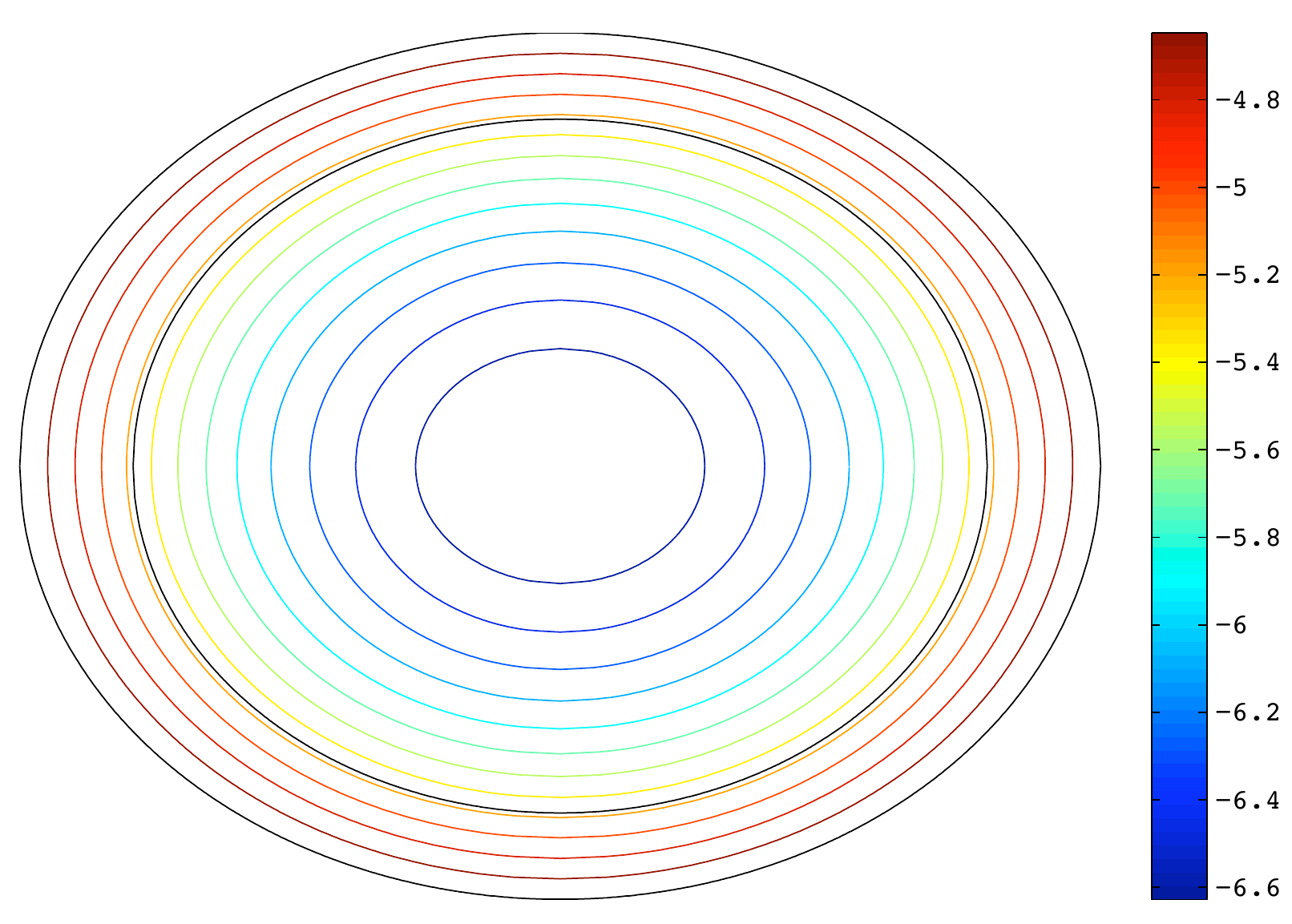}
\includegraphics[scale=.75]{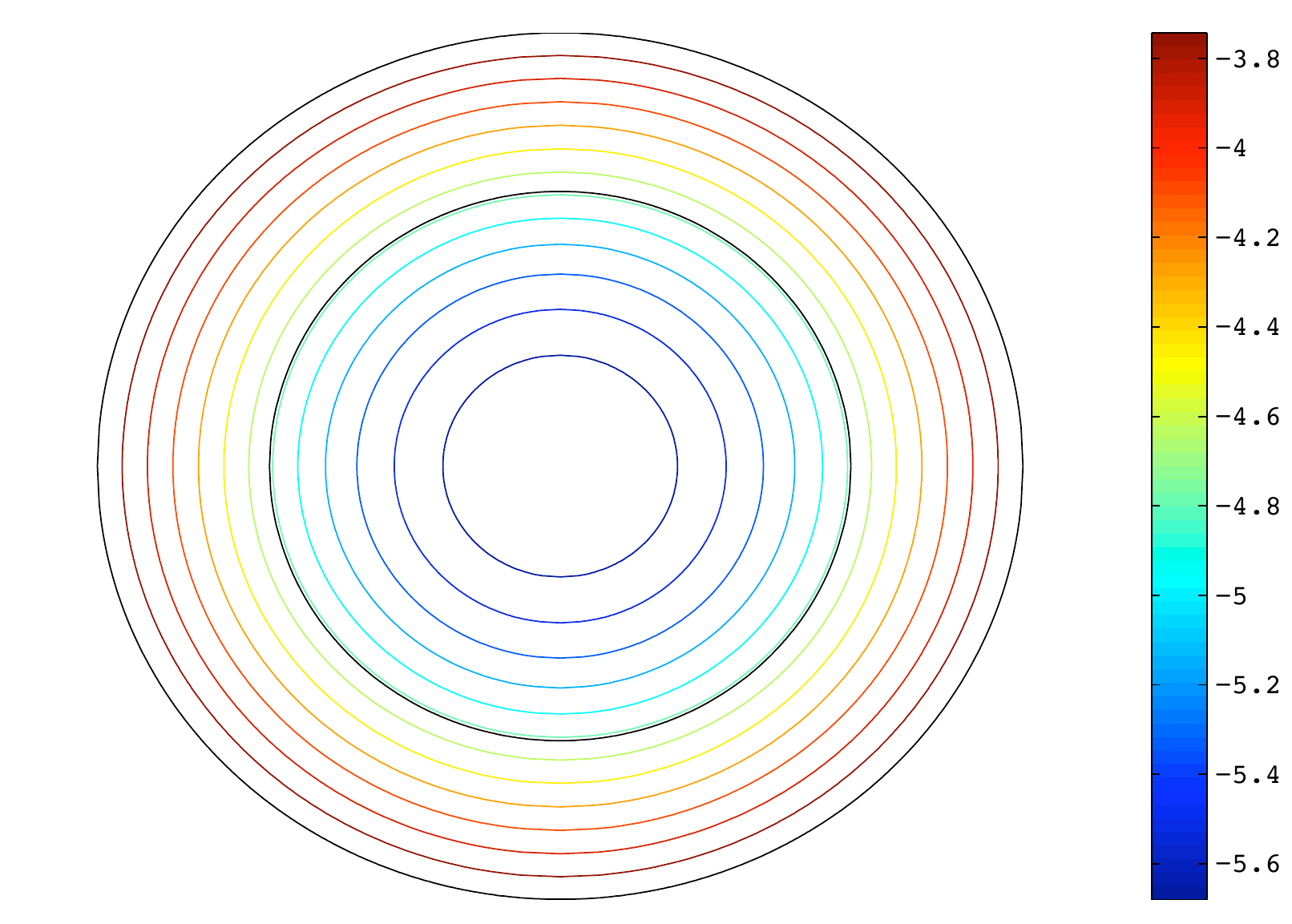}
\caption{Surfaces of constant total (gravity) potential 
 for cases
(a) $Q_{\mathrm{V}} = 0.5$, $\rho_1/\rho_2 = 2$, $\epsilon_2 = 0.18$ and
(b)  $Q_{\mathrm{V}} = 0.25$, $\rho_1/\rho_2 = 2$, $\epsilon_2 = 0.05$.}
\label{fig:1}       
\end{figure}

\begin{figure}
\includegraphics[scale=.8]{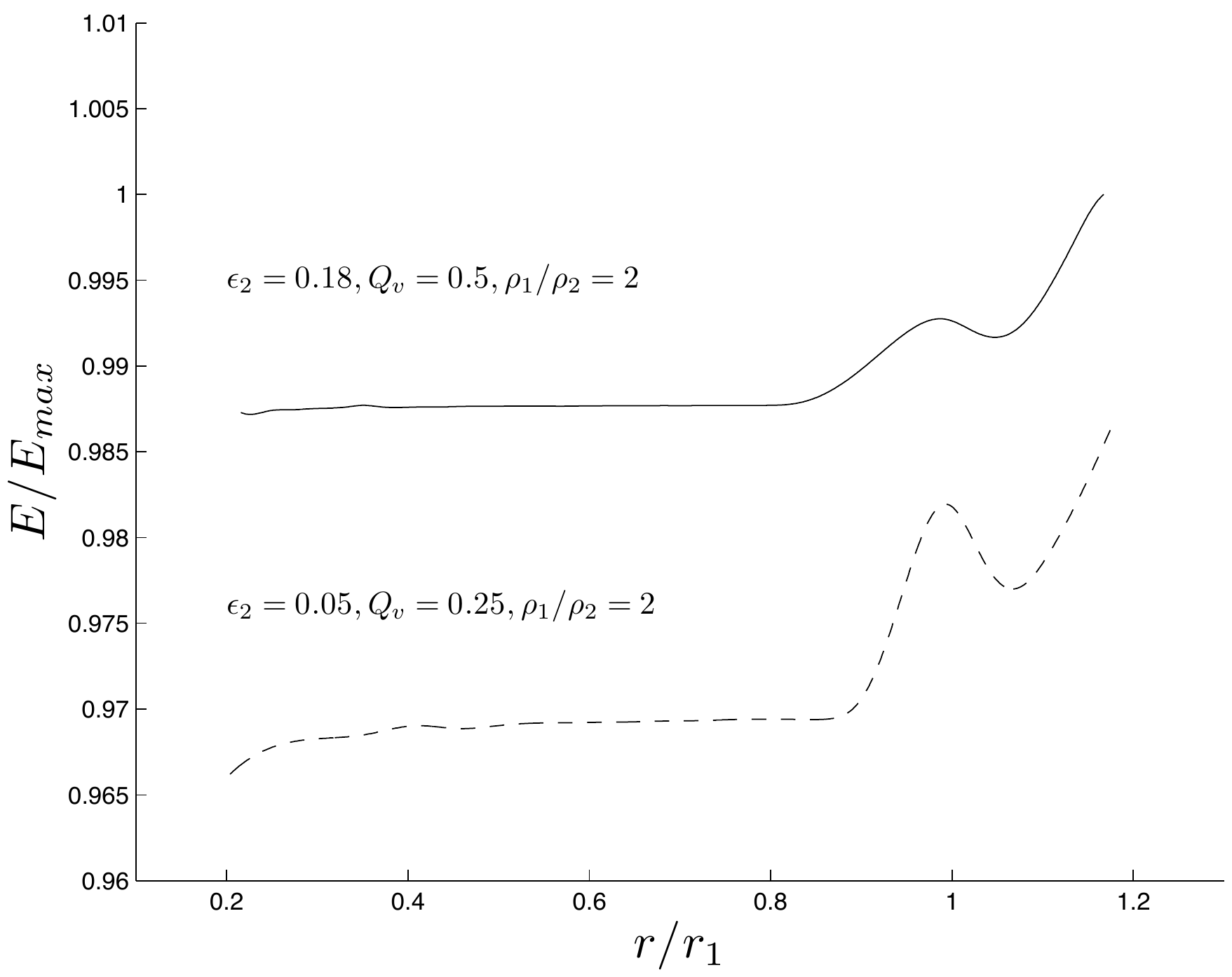}
\caption{The eccentricity of the total potential isosurfaces in Fig.~\protect{\ref{fig:1}} plotted as a
function of radius.}
\label{fig:2}       
\end{figure}

We consider the distortion of a two-layer spherical body rotating with constant angular velocity 
$\Omega$. The core has radius  $r_1$ and constant density  $\rho_1$. The core is surrounded by a spherical shell envelope of outer radius 
$r_2$ and constant density  $\rho_2$. The exact solution for the distortion is presented in
\citet{KongZhangSchub:JGR2010}.
Here, we summarize only a few things necessary for making use of the theory.
The problem is completely specified by three dimensionless parameters, the core-envelope density ratio  
$\rho_1/\rho_2$, the fractional volume of the core  $Q_{\mathrm{V}} = \left(r_1/r_2\right)^3$, and the rotation parameter
\begin{equation}
\epsilon_2 = \frac{\Omega^2}{\left(2\pi \mathrm{G}\rho_2\right)}, 
\label{eq:threepara}
\end{equation}
where G is the gravitational constant.
Among the quantities derivable from the solution are the eccentricities of the
total (gravity) equipotential surfaces, and, in particular, the eccentricities of the core-envelope interface
and the surface, $E_1$ and $E_2$ , respectively.
The total or gravity potential is the sum of the gravitational potential and
the rotational potential.
Surfaces of constant total (gravity) potential are shown in Figure~\ref{fig:1} for the cases
(a) $Q_{\mathrm{V}} = 0.5$, $\rho_1/\rho_2 = 2$, $\epsilon_2 = 0.18$ and
(b)  $Q_{\mathrm{V}} = 0.25$, $\rho_1/\rho_2 = 2$, $\epsilon_2 = 0.05$.
The eccentricity of these total potential isosurfaces is plotted as a
function of radius in Figure~\ref{fig:2}.
The eccentricity of total potential isosurfaces generally decreases inward except for the region
of the interface where eccentricity changes rapidly 
and non-monotonically. Eccentricity has a local maximum at the interface and a local minimum in the envelope
just above the interface. Eccentricity decreases monotonically with decreasing radius inside the core;
the decrease is very gradual until the center of the core is approached.\par

The coefficient $J_2$ of the external gravitational field is an important quantity to determine
because it can be measured by a spacecraft flying by or orbiting a planet.
For the rotating two-layer body, $J_2$ can be found from the exact solution
by proceeding as follows.
In a spherical coordinate system, the axisymmetric gravitational potential
$V_{\mathrm{g}}$ outside a uniformly rotating body can be expanded as
\begin{equation}\label{expan}
V_{\mathrm{g}}(r,\theta)=-\frac{\mathrm{G}M}{r}\left[1-J_2\left(\frac{R_{\mathrm{e}}}{r}\right)^2P_2(\cos{\theta})+\cdots\right],
\end{equation}
where $M$ is the mass of the body,
$P_{2n}$ are the Legendre polynomials, $R_{\mathrm{e}}$ is the equatorial radius of the body, 
$r$ is the radial distance from the center of the body $(r> R_{\mathrm{e}})$, and
$\theta$ is the colatitude of the observation point with respect to the
rotation axis.
On the spherical
surface $r=R_{\mathrm{e}}$, the expansion becomes
\begin{equation}
V_{\mathrm{g}}(r=R_{\mathrm{e}},\theta)=-\frac{\mathrm{G}M}{R_{\mathrm{e}}}\left[1-J_2P_2(\cos{\theta})+\cdots\right]. \label{Vgspherical}
\end{equation}
By determining $V_{\mathrm{g}}(r=R_{\mathrm{e}},\theta)$ from the exact solution, $J_2$ can be
 calculated from the projection of the potential onto the expansion
(\ref{Vgspherical}).

The gravitational potential
expansion in the spheroidal coordinate system
employed in the exact solution is
\begin{eqnarray}
&&V_{\mathrm{g}}(\xi,\eta)=-2\pi {\mathrm{G}} c_2^2\sum_{l=0}^{\infty}i(2l+1)P_l(\eta)  \nonumber \\
& \times & \!\! \Big[
 Q_l(i\xi) \int_0^\xi\int_{-1}^1 \!\!
{[({\xi}^{\prime})^2+({\eta}^{\prime})^2]
P_l({\eta}^{\prime}) P_l(i{\xi}^{\prime})} {\rho}^{\prime}  {\rm d}
{\eta}^{\prime} {\rm d} {\xi}^{\prime}      \nonumber   \\
& + & \!\!  P_l(i\xi) \int_{\xi}^{\xi_{\mathrm{o}}} \!\! \int_{-1}^1
{[({\xi}^{\prime})^2+({\eta}^{\prime})^2]
P_l({\eta}^{\prime})Q_l(i{\xi}^{\prime})} {\rho}^{\prime}  {\rm d}
{\eta}^{\prime} {\rm d} {\xi}^{\prime} \Big]\label{Vgellip}
\end{eqnarray}
\citep{KongZhangSchub:JGR2010}.
In (\ref{Vgellip}), $\xi$, $\eta$ are spheroidal coordinates
with focal length $c_2$, $P_l$ and $Q_l$ are Legendre functions
of the first and second kind, respectively, and $i$ is the
square root of $-1$. In addition $\rho^\prime$ is the density
as a function of  $\xi^\prime$, $\eta^\prime$, and $\xi_{\mathrm{o}}$ is the value of
$\xi$ at the outer free surface.
Evidently, we need to make the transformation from spheroidal
coordinates to spherical coordinates before computing
$V_{\mathrm{g}}(r,\theta)$ from (\ref{Vgellip}). The relationship
between the spherical  and spheroidal coordinates is
\citep{KongZhangSchub:JGR2010}.
\setlength{\arraycolsep}{.03truein}
\begin{eqnarray}
r\cos{\theta} & =&  c_2\sqrt{(1+\xi^2)(1-\eta^2)}, \\
r\sin{\theta} & =&  c_2 \xi \eta
\end{eqnarray}
Taking $r=R_{\mathrm{e}}$ and using the fact that $c_2=R_{\mathrm{e}} E_2$, we find the
transformation in the form
\begin{eqnarray}
\cos{\theta} & =&  E_2\sqrt{(1+\xi^2)(1-\eta^2)},  \label{thetaxieta}  \\
\sin{\theta} & =&  E_2 \xi \eta   \label{thetaxieta2}
\end{eqnarray}
Equations~(\ref{thetaxieta}) and (\ref{thetaxieta2}) enable us to derive $\xi$ and $\eta$ as functions of
$\theta$
\begin{eqnarray}
\eta^2&=&\frac{\sqrt{(1-E_2^2)^2+4E_2^2\cos^2{\theta}}-(1-E_2^2)}{2E_2^2} \label{etatheta}\\
\xi^2&=&\eta^2+\frac{1}{E_2^2}-1 \label{xitheta}
\end{eqnarray}
With (\ref{etatheta}) and (\ref{xitheta}), we are
able to derive the gravitational potential as a function of the
spherical coordinate $\theta$.

When $V_{\mathrm{g}}(r=R_{\mathrm{e}},\theta)$ is available, we project it onto the
spherical harmonic expansion to obtain $J_2$. We  expect the
expansion in the form
\begin{equation}
V_{\mathrm{g}} \left(r = R_{\mathrm{e}}, \theta\right)=\sum_l{\sqrt{\frac{2l+1}{2}}C_lP_l(\cos{\theta})}, \label{Vgabstract}
\end{equation}
where
\begin{equation}
C_l=\sqrt{\frac{2l+1}{2}}\int_0^\pi{ V_{\mathrm{g}}\left( r= R_{\mathrm{e}}, \theta\right)P_l(\cos{\theta})\sin{\theta}}{\rm d}\theta.
\end{equation}
By comparing (\ref{Vgabstract}) and
(\ref{Vgspherical}), we find
\begin{eqnarray}
\frac{1}{2}C_0&=&-\frac{\mathrm{G}M}{R_{\mathrm{e}}}\\
\frac{5}{2}C_2&=&\frac{\mathrm{G}M}{R_{\mathrm{e}}}J_2,
\end{eqnarray}
from which $J_2$ can be simply calculated as
\begin{equation}
J_2=\frac{-5C_2}{C_0}.
\end{equation}

\section{Comparison with the Radau-Darwin Approximation}
\label{sec:comprd}
The Radau-Darwin approximate formula can be used to predict the flattening or eccentricity
of the outer surface of a rotating body
\citep{Radau:CR1885,Darwi:MNRAS1900}. 
The formula relates the normalized axial moment of inertia 
$C/Ma_2^2$ ($C$ is the axial moment of inertia around the rotation axis,
$M$ is the total mass of the body, and $a_2$  is the equatorial radius) to the
second degree Love number  $h_2$
\begin{equation}
\frac{C}{Ma_2^2} = \frac{2}{3} \left\lbrack 1 - \frac{2}{5}
                                 \frac{\left(5 - h_2\right)^{1/2}}{h_2^{1/2}}\right\rbrack
\label{eq:rd}
\end{equation}
\citep{ZharkTrubi:PPI1978}.
The Love number gives the flattening of the surface $f_2 = \left(a_2 - c_2\right)/a_2$ by
\begin{equation}
f_2 = \frac{qh_2}{2}
\label{eq:loveflatt}
\end{equation}
\citep{ZharkTrubi:PPI1978},
where $q$ is the small rotational parameter
\begin{equation}
q = \frac{\Omega^2a_2^3}{\mathrm{G}M}
\label{eq:rot}
\end{equation}
The flattening and eccentricity of the surface of the body are related by
\begin{equation}
f_2 = 1 - \left( 1 -E_2^2\right)^{1/2}
\label{eq:flattecc}
\end{equation}

Equations~(\ref{eq:rd})-\ref{eq:flattecc})  are often used in planetary physics to determine the moment of inertia of a body
whose shape (flattening), rotation rate, mass, and equatorial radius are known.
With $f_2$ and $q$ known, (\ref{eq:loveflatt}) gives $h_2$ and (\ref{eq:rd}) gives 
$C/Ma_2^2$.  Alternative to the flattening, the gravitational coefficient $J_2$ can be used to infer the axial moment of
inertia since to a first approximation $J_2$ and $f_2$ are related by
\begin{equation}
f_2 = \frac{3}{2} J_2 + \frac{1}{2} m
\label{eq:jf}
\end{equation}
\citep{ZharkTrubi:PPI1978},
where $m$ is the small rotational parameter given by
\begin{equation}
 m = \frac{\Omega^2s_2^3}{\mathrm{G}M}
 \label{eq:mdef}
\end{equation}
and $s_2$ is the mean radius of the body (the radius
of a spherical body with the same volume as the
rotationally-distorted body).
The Radau-Darwin formula can be rewritten
in terms of $J_2$ and $m$ by using (\ref{eq:loveflatt}) and (\ref{eq:mdef}) to eliminate $h_2$.\par

The Radau-Darwin formula can be used to predict the flattening or eccentricity of the surface
of the rotating two-layer spherical body and the result compared with the value
of the eccentricity from the exact solution of
\citet{KongZhangSchub:JGR2010}.
The normalized moment of inertia of a two-layer sphere is
\begin{equation}
\frac{C}{Mr_2^2} = \frac{2}{5} \left\lbrack
 \frac{\rho_1}{\overline\rho} +
  \frac{\left(\rho_1 - \rho_2\right)}{\overline\rho}
   \left(\frac{r_1}{r_2}\right)^5 \right\rbrack
\label{eq:nmominert}
\end{equation}	
This can be rewritten in terms of the dimensionless variables that characterize the
solution of 
\citet{KongZhangSchub:JGR2010}
as
\begin{equation}
\frac{C}{Mr_2^2} = \frac{2}{5} \left\lbrack \left(1 + Q_{\mathrm{V}}^{5/3}\right)
          \left\lbrace Q_{\mathrm{V}} + \frac{\left(1-Q_{\mathrm{V}}\right)}{\rho_1/\rho_2}\right\rbrace^{-1}
          - Q_{\mathrm{V}}^{5/3}\left\lbrace 1 - Q_{\mathrm{V}} + \left(\frac{\rho_1}{\rho_2}\right)
           Q_{\mathrm{V}}\right\rbrace^{-1}\right\rbrack
\label{eq:dvar}
\end{equation}	
Given $Q_{\mathrm{V}}$ and $\rho_1/\rho_2$, the dimensionless moment of inertia is
calculated from (\ref{eq:dvar}), and $h_2$ follows from (\ref{eq:rd}).
The flattening and eccentricity of the model surface is then
obtained from (\ref{eq:loveflatt}) and (\ref{eq:flattecc}).\par

Table~\ref{tab:xy} compares the exact solution for the eccentricities of the interface $E_1$ and surface 
$E_2$  (evaluated using the theory of
\citet{KongZhangSchub:JGR2010}) with the eccentricity of the surface from the Radau-Darwin formula $E_2^{R-D}$ for the case
$Q_{\mathrm{V}} = 0.5$ and $\rho_1/\rho_2 = 2$ for different values of the rotation parameter $\epsilon_2$.
The agreement between the approximate and exact solutions is quite good in this case for all the rotation
rates considered. Strictly speaking, the Radau-Darwin approximation can be said to be invalid for values of
$\epsilon_2 > 0.17$ since the Radau-Darwin formula predicts surface eccentricities less than the
interface eccentricities, when in fact, the interface eccentricity should be less than the surface eccentricity.
That this is the case can be qualitatively understood by considering the flattening of a rotating
sphere of uniform density $\rho$. The flattening is proportional to $\rho^{-1}$.
In a body that has density increasing with depth, the flattening or eccentricity of equipotential
surfaces should accordingly decrease with depth.\par

\section{Comparison with the Theory of Figures for the Generalized Roche Model}
\label{sec:comp}

The generalized Roche model is a special case of the two-layer model of this paper in which the envelope density
$\rho_2 = 0$ . An analytic formula for the flattening of total potential isosurfaces, correct to second order
in the rotational parameter $m$, is given in
\citet{ZharkTrubi:PPI1978}.
(We also derive this formula later in Section~5.8.1,
which discusses the generalized Roche model from our perspective
on the theory of figures.)
The flattening of the surface according to this formula $f_2^{\mathrm{ToF}}$ is given by
\begin{equation}
 f_2^{\mathrm{ToF}} = \frac{1}{2}m \left(1 + \frac{3}{2} \beta_C^5\right)
                       + \frac{1}{8} m^2 \left(\beta_C^5 + \frac{20}{7}\beta_C^8
                        - \frac{33}{28}\beta_C^{10}\right)
 \label{eq:surfflat}
\end{equation}		 		
where $\beta_C = r_1/r_2 = Q_{\mathrm{V}}^{1/3}$. The rotation parameter $m$ is defined in
(\ref{eq:mdef})for $s_2 = r_2$. It is related to the parameters of the exact theory by
\begin{equation}
 m = \frac{3\left(\epsilon_2\rho_2\right)}{2Q_{\mathrm{V}}\rho_1}
 \label{eq:rottheory}
\end{equation}
where $\epsilon_2\rho_2$  is independent of  $\rho_2$
(see (\ref{eq:threepara})).
The flattening of the interface according to the second order theory
of figures ToF is
\begin{equation}
  f_1^{\mathrm{ToF}} = \frac{5}{4}m\beta_c^3 + \frac{75}{224} m^2\beta_c^6
  \label{eq:f1tof}
\end{equation}
Equation~(\ref{eq:f1tof}) follows from  (\ref{FKH}) and (\ref{SolF1Roche}) and (\ref{SolF2Roche})
with $\beta = \beta_c$ in the expressions for $F_1$ and $F_2$.

Table~\ref{tab:yz} compares the eccentricity of the surface and the interface of several models computed
using the exact theory with values obtained from (\ref{eq:surfflat}) and (\ref{eq:f1tof})
 for the generalized Roche model
with the help of (\ref{eq:flattecc}) to convert  to the eccentricity $E_2^{\mathrm{ToF}}$ .
For the exact theory we consider the sequence of values of 
$\rho_2/\rho_1$ equal to $10^{-2}$, $10^{-3}$, and $10^{-4}$ to compare with
the Roche model result that takes $\rho_2 = 0$. For the cases considered
in the table the theory of figures to second order in $m$ only slightly underestimates
the surface and interface eccentricities.

A formula for the surface flattening correct to third order in $m$
 is given by combining $f\left(\beta_c\right)$ from
 (\ref{eq:f1tof}) with
 $F_{31}$ given by
 (\ref{H31K31F31Roche}).
 
\section{First-Order Theory of Figures for Synchronous Rotation and Tides}      

The theory of figures dates back to Clairaut, who in 1743 derived an integrodifferential equation
for the flattening of a rotating body in hydrostatic equilibrium (HE), but with a non-uniform density distribution in the interior 
\citep{Kaula:IPPTP1968}.
Clairaut's theory represents a first order perturbation theory to a non-rotating spherical configuration
with arbitrary density distribution in layers, the level surfaces, or the surfaces of constant total potential,
the sum of the gravitational potential and the non-inertial centrifugal potential
(zero in the spherical configuration). The small rotational parameter $m$ for the perturbation theory
is related to the body's rotation period $P$, its mean radius $R$, and its total mass $M$ by
\begin{equation}
  m =   \left( \frac{2 \pi}{P} \right)^2 \left( \frac{R^3}{\mathrm{G}M} \right)\quad (\mathrm{compare~(\ref{eq:mdef})})
\label{m}
\end{equation} 
where G is the gravitational constant given by $(6.674215 \pm 0.000092) \times 10^{-11}~\mathrm{m}^3~\mathrm{kg}^{-1}~\mathrm{s}^{-2}$
 \citep{GundlMerko:PRL2000}.
Often, the combination G$M$ is known from orbital dynamics to greater accuracy than
G itself, and the total mass is a derived parameter given by G$M$/G, accurate to essentially the same
 fractional error as G, or 14 parts per million. The mean radius  $R$ (also denoted by
 $s_2$ in (\ref{eq:mdef})) is the radius of a uniform sphere
 equal in density to the planet's mean density $ \rho_0$ or,
\begin{equation}
  \rho_0 = \frac{3 \left(\mathrm{G}M \right)}{4 \pi \mathrm{G} R^3} 
\label{rho0}
\end{equation}
The flattening $f$ is defined in terms of the equatorial radius $a$ and polar radius $c$ by,
\begin{equation}
  f =   \frac{a - c}{a}
\label{f}
\end{equation}
In the first order Clairaut theory, the rotating mass configuration consists of continuous layers of
concentric ellipsoids of revolution, each with its own density and flattening.
The configuration is defined in terms of a single variable, the normalized mean radius
$ \beta = s/R$, which labels the level surfaces. The actual mean radius $s$, in metric units,
can always be recovered by multiplying $\beta$ by the given mean radius $R$ of the planet.
The density $\rho(\beta)$ of a particular level surface can also be normalized to the
given mean density, so that $ \delta(\beta) = \rho(\beta)/\rho_0$.\par

For perturbations of higher order than the first, these definitions can all be retained.
However, the level surfaces are no longer ellipsoids of revolution. Even so, the radius
$r$ of a particular level surface can be expressed in terms of the mean radius $s$ by a distortion to the
usual polar coordinate equation for an  ellipse. For the third order theory, the radius $r$ can be expressed as a
function of the polar angle $\theta$ (colatitude), the flattening $f$, and two higher-order,
spheroidal shape parameters $k$ and $h$, as follows
 \citep{ZharkTrubi:PPI1978}.
\begin{equation}
  r \left( \theta \right) =   a \left[ 1 - f \cos^2 \theta  - \left( \frac{3}{8} f^2 + k \right) \sin^2 2 \theta
+\frac{1}{4} \left( \frac{1}{2} f^3 + h \right) \left( 1 - 5 \sin^2 \theta  \right) \sin^2 2 \theta \right]
\label{r}
\end{equation}

It is convenient to express the radius in terms of $\mu$, the cosine of $\theta$.
Then the mean radius is given by the integral
\begin{equation}
  s^3 =   \frac{1}{2} \int^{1}_{-1}r^3 \left( \mu \right)\mathrm{d} \mu
\label{s}
\end{equation}
and the expression for the radius becomes
\begin{equation}
  r =   R \beta \sum^{\infty}_{i=0} a_{2 i} \mu^{2 i} 
\label{rmu}
\end{equation}
Consistent with (\ref{r}) and (\ref{s}), the coefficients to third order are
\begin{eqnarray}
  a_0 &=&   \left( 1+ \frac{1}{3} f + \frac{2}{9} f^2 + \frac{14}{81} f^3  + \frac{8}{15} k +  \frac{26}{105} h + \frac{16}{63} f k \right) \nonumber \\
  a_2 &=&   - \left( f + \frac{11}{6} f^2 + \frac{49}{18} f^3 + 4 k + 4 h + \frac{28}{15} fk \right) \nonumber \\
  a_4 &=&   \frac{3}{2} \left(f^2 + \frac{10}{3} f^3 + \frac{8}{3} k + 6 h  + \frac{8}{9} f k \right) \nonumber \\
  a_6 &=&   - \frac{5}{2} \left( f^3 + 2 h \right)
\label{a2i}
\end{eqnarray}
This is an alternating series that converges absolutely, and the error in the partial sum is less
 than the absolute value of the next term in the series.

Equation~(\ref{rmu}) is fundamental to the theory of figures. With it, a coordinate transformation between Cartesian
coordinates ($x$,$y$,$z$) and generalized coordinates $ (\beta, \phi, \mu)$ can be defined by
\begin{eqnarray}
  x &=&   r \cos \phi \sqrt{1 - \mu^2} \nonumber \\
  y &=&   r \sin \phi \sqrt{1 - \mu^2} \nonumber \\
  z &=&   r \mu 
\label{coord}
\end{eqnarray} 
Because of axial symmetry, the azimuthal coordinate $\phi$ can be integrated out of the problem immediately.
The series coefficients $a_{2 i}$ are functions of $\beta$ only, by means of the shape functions {$f$,$k$,$h$}.
The differential volume element d$\tau$ can be found from the Jacobian determinant of the transformation.

For example, after integrating for the coordinates $\phi$ and $\mu$, the gravitational coefficients
$J_n$ in the external gravitational potential are given by
\citet{ZharkTrubi:PPI1978},
\begin{equation}
  J_n = - \int \delta \left( \beta \right) \beta^n P_n \left( \mu \right)\,\mathrm{d} \tau  =   - \int^1_0 \delta \left( \beta \right)\,\mathrm{d} \left[ \beta^{n+3} \phi_n \left( \beta \right) \right],
\label{Jn}
\end{equation}
where $ P_n$ is the Legendre polynomial of degree $n$.
The functions $ \phi_n$ are derived in \ref{AppendixA} to third order by means of a definite integral in
$\mu$ from minus one to plus one. The only integral that remains in (\ref{Jn}) for further evaluation is the integral in
$\beta$, which depends on the given density distribution $\delta (\beta)$.

Similarly, the principal moments of inertia, $C$ along the polar axis, and $A$ along an equatorial axis, can be expressed
as third-order series. All quantities are thereby normalized to the total mass $M$ and powers of the mean radius $R$,
so that the external gravitational potential function $V$ is expressed as
\begin{equation}
  V =   \frac{\mathrm{G} M}{r} \left[1 - \sum^{\infty}_{i=1}  \left( \frac{R}{r}  \right)^{2 i} J_{2 i} P_{2 i} \left( \mu \right)    \right] 
\label{V}
\end{equation}
The measured gravitational coefficients $J_{2 i}$ can be determined from orbital dynamics, as is G$M$.
They are most often referred to a reference value for the equatorial radius $a_{\mathrm{ref}}$. However, they can be referred to $R$
by multiplying each observed value of degree $2i$ by $ (a_{\mathrm{ref}}/R)^{2i}$, consistent with the computed values from (\ref{Jn}).

The usefulness of the theory of figures is that a reasonable density distribution $\delta(\beta)$ can be found that minimizes the
weighted sum of squares WSOS for the measured coefficients, the method of weighted least squares. The minimization function in terms of the
observed coefficients $  \hat{J}_{2i}$, along with their standard errors $ \sigma_{2 i}$, and the computed values $  J_{2i}$ from (\ref{Jn}) is
\begin{equation}
  WSOS =   \sum^{\infty}_{i=1} \left( \frac{\hat{J}_{2i}-J_{2 i}}{\sigma_{2 i}} \right)^2 
\label{WSOS}
\end{equation}
If the orbital dynamics is limited, such that there are strong correlations among the coefficients, the minimization can be generalized
to include the covariance matrix $\Gamma$ for the $ \hat{J}_{2i}$ from the analysis of the orbital data. The residuals $ \hat{J}_{2i}-J_{2 i}$
are placed in a column matrix z and WSOS is defined by the matrix operation $ z^T \Gamma^{-1} z$, with $  z^T$ the transpose of $z$.

In principle the theory can be extended to arbitrary order in the small rotational parameter $m$. However, it becomes quite cumbersome for orders
greater than three. Nevertheless, all expressions required for a fifth-order theory have been published by
\citet{ZharkTrubi:SA1976}.
 In the fifth-order theory the radius $r$ is still normalized to the mean radius $s$, but (\ref{rmu}) is replaced by the following Legendre series, valid to arbitrary order $n$
\begin{equation}
  \frac{r}{s} =   1 + \sum_{j=2}^n s_{0 j} m^j + \sum_{j=1}^n m^j s_{2 j} P_{2 j} \left( \mu \right)
\label{rLP}
\end{equation}
This expression can be substituted into  (\ref{s}) and expanded in a power series in $m$ to order $n$.
Then coefficients of $m^i$ for powers of $i$ greater
than or equal to 2 can be set to zero. This yields $n-1$ equations in the $n-1$  coefficients $s _{0 j}$, which can
 all be evaluated. For example, at fifth order the values are
\begin{eqnarray}
  s_{0 2} &=&   - \frac{1}{5} s_2^2 \nonumber \\
  s_{0 3} &=&   - \frac{2}{105} s_2^3 \nonumber \\
  s_{0 4} &=&   - \frac{1}{315} \left( 35 s_4^2 + 18 s_2^2 s_4 \right) \nonumber \\
  s_{0 5} &=&   - \frac{2}{17325} \left( 33 s_2^5 + 250 s_2 s_4^2 \right) \label{s0Fcns}
\end{eqnarray}
A substitution of the $s _{0 j}$ values so determined into (\ref{rLP}) yields the $n^{th}$ order expression for $r/s$,
comparable to (\ref{rmu}), the third-order expression in the spheroidal functions $f$, $k$, and $h$.
The parameter $s_2$ in (\ref{rLP}) and (\ref{s0Fcns}) is not the mean radius parameter used in (\ref{eq:mdef}).

The easiest way to compare the theory with observation is to refer the measured gravitational harmonics to the mean
radius of the planet. The advantage is that (\ref{Jn}) directly represents what is being measured. However, the mean
 radius of a planet depends on its rotation period, which is not always known, and which may not even be constant throughout the interior.
 In this sense, the measured equatorial radius is a more fundamental observational constraint. Therefore, when the measured
 $J _n$ are referred to the equatorial radius, the theoretical values given by (\ref{Jn}) must be multiplied by the ratio $(s/a)^n$.
 This ratio can be found from the inverse of the expression for $r/s$, with $\mu$ set equal to zero. With this approach, care is required in
 order to make sure that higher order terms in $(s/a)^n$ do not enter into the theoretical expression for $J _n$ and bias it.
 Furthermore, this is true in general. If one is computing a first-order Clairaut spheroid, it is important to make sure 
 the series are always
 strictly truncated at first order. The same can be said for a second-order Darwin spheroid, or to any spheroid of arbitrary order $n$.

\subsection{The Level Surfaces}
The application of (\ref{Jn}) requires that the functions $s _{2 j}(\beta)$, or alternatively $f(\beta)$, $h(\beta)$, and $k(\beta)$
in the expression for $r/s$ of  (\ref{rmu}), be found for a given interior density distribution, expressed in terms of the normalized
density $\delta(\beta)$. This is accomplished by finding the level surfaces on which the interior gravitational potential is a
constant. The interior potential at a normalized mean radius $\beta$ is determined by the amount of mass interior to the level surface,
an integral over the volume from zero to $\beta$, plus the amount of mass exterior to the level surface, an integral from $\beta$ to one.
Let the gravitational moments of the mass  lying internal to $\beta$ be given by $S _{2 j}(\beta)$. The normalized potential $V _0$ from the
mass interior to $\beta$ is given by the expansion of $s/r$ in Legendre polynomials. We illustrate this procedure for the spheroidal
functions to order three. The more general procedure for the $s _{2 j}$ functions to arbitrary order is similar. 

The first step is to invert the expression for $r/s$ in (\ref{rmu}) and expand it in a series to order three in the small
rotational parameter $m$.
The result is a function in even powers of $\mu$. Next, the Legendre polynomials in $\mu$ can be inverted to any
arbitrary degree to obtain powers of $\mu$ in terms of the polynomials. For third order in $m$ the result is
\begin{eqnarray}
  \mu^2 &=&   \frac{1}{3} \left[ 1 + 2 P_2 \left( \mu \right) \right] \nonumber \\
  \mu^4 &=&   \frac{1}{35} \left[ 7 + 20 P_2 \left( \mu \right) + 8 P_4 \left( \mu \right)\right] \nonumber \\
  \mu^6 &=&   \frac{1}{231} \left[ 33 + 110 P_2 \left( \mu \right) +  72 P_4 \left( \mu \right) + 16 P_6 \left( \mu \right) \right]
\label{mun}
\end{eqnarray}
The next step in the procedure is to substitute the powers of $\mu$ given by  (\ref{mun}) into the series for $s/r$.
The result is an expansion of $s/r$ in a series of Legendre polynomials in the form
\begin{equation}
  V_0 = \frac{s}{r} =   \sum_{j=0}^n C_{2 j}^0 P_{2 j} \left( \mu \right)
\label{sbyr}
\end{equation}
For the spheroidal functions, the coefficients $C _{2 j}^0$ can be written as
\begin{eqnarray}
  C_0^0 &=&   1 + \frac{8}{45} f^2 + \frac{584}{2835} f^3 + \frac{64}{315} f k \nonumber \\
  C_2^0 &=&   \frac{2}{3} f + \frac{31}{63} f^2 + \frac{76}{189} f^3 - \frac{2}{21} h + \frac{8}{21} k + \frac{88}{315} f k \nonumber \\ 
  C_4^0 &=&   - \frac{4}{35} f^2 - \frac{172}{1155} f^3 - \frac{192}{385} h - \frac{32}{35} k - \frac{416}{1155} f k \nonumber \\
  C_6^0 &=&   \frac{8}{231} f^3 + \frac{80}{231} h - \frac{128}{231} f k
\label{C2j0}
\end{eqnarray}
This completes the expansion for the zero-degree gravitational moment $S _0$,
which is basically a mass function given by
\begin{equation}
  S_0 =   \frac{3}{\beta^3} \int_0^\beta z^2 \delta \left( z \right) \, \mathrm{d} z 
\label{S0}
\end{equation}
For any interior density distribution given by $\delta(\beta)$, the function $S _0$ must be equal to one at the surface of the
planet, where $\beta$ is equal to one. Otherwise the interior model will not be consistent with the observed mass and mean radius.

In general, the gravitational moments $S _{2i}$ are included in the level-surface theory by series expansion in powers
of the inverted $r/s$ in (\ref{sbyr}) times the appropriate higher-degree Legendre polynomial, as follows
\begin{equation}
  V_i = \left( \frac{s}{r} \right)^{2 i + 1} P_{2 i} \left( \mu \right)~~~~~i = 0,1,2, \cdot \cdot \cdot, n
\label{VS}
\end{equation}
The series expansion to order $m$ for a particular degree $2i$ is carried out to order $n-i$.  The powers of $\mu$ given by (\ref{mun})
are substituted into the series for $V _i$. The result is an expansion in Legendre polynomials that can be written 
\begin{equation}
  V_i =   \sum_{j=0}^n C_{2j}^i P_{2 j} \left( \mu \right)~~~~~i = 0,1,2, \cdot \cdot \cdot, n
\label{VSP}
\end{equation}
The evaluation of the coefficients $C_{2j}^i$ for the spheroidal functions is given in
\ref{AppendixA} for orders 1, 2 and 3. The coefficients for order zero are given by (\ref{C2j0}).

The gravitational moments $S _{2i}^\prime$ for the potential exterior to the level surface labeled by
$\beta$ require potential functions  $V _i^\prime$, which are defined by
\begin{equation}
  V_i^\prime = \left( \frac{r}{s} \right)^{2 i} P_{2 i} \left( \mu \right)~~~~~i = 0,1,2, \cdot \cdot \cdot, n
\label{VSp}
\end{equation}
After similar expansion in powers of $m$ as for $V _i$, the potentials for mass between $\beta$
and the surface at $\beta = 1$ can be expressed in terms of coefficients $ C_{2j}^{i \prime}$ by
\begin{equation}
  V_i^\prime =   \sum_{j=0}^n C_{2j}^{i \prime} P_{2 j} \left( \mu \right)~~~~~i = 0,1,2, \cdot \cdot \cdot, n
\label{VSPp}
\end{equation}
The coefficients $C_{2j}^{i \prime}$ are given to third order in \ref{AppendixA}.

So far we have been concerned with the expansion of the internal gravitational potential to order $n$ in a series of
Legendre polynomials of degree $2n$ in the general coordinate $\mu$. However, the planet is in rotation about its principal axis of maximum
moment of inertia, the $z$ axis. In this rotating non-inertial coordinate system the planet deviates from a sphere because of a centrifugal force
generated  by a rotation in inertial space, a rotation with respect to the ``fixed stars''. Relativistic corrections to this Newtonian model are
ignored in the theory of figures for planets. Therefore the centrifugal force per unit mass can be represented by the following potential function $V_{\mathrm{rot}}$
\begin{equation}
  V_{\mathrm{rot}} = \frac{1}{2} \left( \frac{2 \pi}{P} \right)^2 r^2 \sin^2 \theta
\label{Q}
\end{equation}
This potential function can be made consistent with the gravitational potentials $V _i$ and $V _i^\prime$ by replacing the period $P$
by the smallness parameter $m$ according to (\ref{m}), by replacing $\mu$ by the Legendre polynomial $P _2$ according to 
(\ref{mun}), and by normalizing to the gravitational potential G$M/R$ at the surface. The result is
\citep{ZharkTrubi:PPI1978}
\begin{equation}
  Q =   \frac{1}{3} m \left( \frac{r}{s} \right)^2 \left[ 1 - P_2 \left( \mu \right) \right]
\label{Qm}
\end{equation}
The centrifugal potential $Q$ can be expanded to arbitrary order by means of  (\ref{rLP}) or to third order by 
(\ref{rmu}).
The third order coefficients corresponding to the third order coefficients for  $V _i$ and $V _i^\prime$ are
\begin{eqnarray}
  Q_0 &=&   m \left( \frac{1}{3} + \frac{4}{45} f + \frac{2}{189} f^2 + \frac{16}{315} k \right) \nonumber \\
  Q_2 &=&   - m \left( \frac{1}{3} +\frac{20}{63} f + \frac{38}{189} f^2 + \frac{16}{45} k \right) \nonumber \\ 
  Q_4 &=&   m \left( \frac{8}{35} f + \frac{76}{231} f^2 + \frac{32}{55} k \right) \nonumber \\
  Q_6 &=&   - m \left( \frac{32}{231} f^2 + \frac{64}{231} k \right)
\label{Qj}
\end{eqnarray}
The small rotational  parameter $m$ enters explicitly in the theory of figures by means of the centrifugal potential $Q$. 

All the coefficients derived so far can be collected into complete expressions for the internal potentials
$A _{i}$ on the level surface labeled by $\beta$. These potentials to arbitrary order can be written as
\begin{equation}
  A_{i} =   \sum_{j=0}^n \left( C_{2 j}^{i} S_{2 j} +  C_{2 j}^{i \prime} S_{2 j}^\prime \right) + Q_i~~~~~i = 0,2,4, \cdot \cdot \cdot, 2 n
\label{A2i}
\end{equation}
The gravitational moments can be written in terms of the following integrals
 \citep{ZharkTrubi:PPI1978}
\begin{eqnarray}
  S_{i} &=&   \frac{1}{\beta^{i + 3}} \int_0^\beta \delta \left( z \right)\,\mathrm{d} \left[ z^{i+3} \phi_i \right] \nonumber \\
  S_{i}^\prime &=&   \beta^{i-2} \int_\beta^1 \delta \left( z \right) \, \mathrm{d} \left[ z^{2-i} \phi_i^\prime \right]
\label{Sn}
\end{eqnarray}
The functions $  \phi_i$ and $  \phi_i^\prime$ represent the integral of the gravitational moments over $\mu$ as follows
\begin{eqnarray}
  \phi_i &=&   \frac{3}{2 \left( i + 3 \right)} \int_{-1}^{1} P_i \left( \mu \right) \left( \frac{r}{s} \right)^{i+3}\, \mathrm{d} \mu~~~~~i = 0,2,4, \cdot \cdot \cdot, 2 n \nonumber \\
  \phi_i^\prime &=&   \frac{3}{2 \left( 2-i \right)} \int_{-1}^{1} P_i \left( \mu \right) \left( \frac{r}{s} \right)^{2-i} \, \mathrm{d} \mu~~~~~i = 0,2,4, \cdot \cdot \cdot, 2 n
\label{phi}
\end{eqnarray}
When $i$ is equal to 2, the integration for $\phi_2^\prime$ must be carried out as a special limiting case. The integral is
\begin{equation}
  \phi_2^\prime =   \frac{3}{2} \int_{-1}^1 P_2 \left( \mu \right) \ln \left( \frac{r}{s} \right)\,\mathrm{d} \mu
\label{phi2}
\end{equation}
The functions under the integrals for $ \phi_i$ and $  \phi_i^\prime$ can be expanded to arbitrary order in $m$ and integrated.
The results to order three are given in
\ref{AppendixA}. Results to order 5 by means of  (\ref{rLP}) are given by
\citet{ZharkTrubi:SA1976}.

Evaluations of $A _0$,  $A _2$,  $A _4$ and  $A _6$ are given in 
Appendix~B. 
In order that the potential be a constant on
level surfaces, all potentials of order greater than zero must be zero. This means that any $A _i$ with $i$ equal to 2 or greater can be multiplied
through by a constant. It also means that $A _2$ can be used to solve for $m$, as it appears explicitly in $Q_2$. By substituting this value of $m$
into the higher-degree potentials, $A _4$, $A _6$ and higher, they can be simplified. They do not contain $m$ explicitly. $A _0$ is the only potential
 function that is not zero. For this reason it represents the total internal potential at normalized mean radius $\beta$, with the centrifugal term included in the potential.
 It is the potential that enters in the equation of HE. The pressure $p(\beta)$ on a level surface and the total gravitational potential $U(\beta)$ can be normalized
 by the following relations involving the given mass $M$ and mean radius $R$ for the planet
\begin{eqnarray}
  \chi \left( \beta \right) &=&   \frac{R p \left ( \beta \right)}{\mathrm{G}M \rho_0} \nonumber \\
  A_0 \left( \beta \right) &=&   \frac{R U \left( \beta \right)}{\mathrm{G}M}
\label{pUNorm}
\end{eqnarray}
In terms of these normalized variables, the equation of HE in the interior is given by
\citet{ZharkTrubi:PPI1978}
\begin{equation}
  \frac{\mathrm{d } \chi}{\mathrm{d}\beta} =   \delta \frac{\mathrm{d}\left( \beta^2 A_0 \right)}{\mathrm{d}\beta}
\label{HE}
\end{equation}

\subsection{Solution to the Level-Surface Problem}
\label{Sol}
The objective of a solution to the level-surface problem is to find the gravitational moments and the shape of the planet at its surface,
and to compare the theoretical result with what is observed for the surface shape and external gravitational field.
This result depends on the interior normalized density distribution $\delta ( \beta )$, which can be a given function, as in the two-zone model treated here,
or it can be obtained from a known equation of state (EOS) by including (\ref{HE}) in the solution of the overall problem.
When the EOS  is given for the internal material as a function $\chi(\delta)$, most likely in zones, the following
differential equation for $\delta$ can be included in the solution for the theoretical interior model
\begin{equation}
  \left( \frac{\mathrm{d}\chi}{\mathrm{d}\delta} \right) \frac{\mathrm{d} \delta}{\mathrm{d}\beta}
  =   \delta \frac{\mathrm{d} \left( \beta^2 A_0 \right)}{\mathrm{d}\beta} 
\label{HEI}
\end{equation}
This suggests that it might be useful to have the level-surface problem not in the form of integrodifferential equations, but in the form of
differential equations only. An advantage of this approach is that a numerical solution to a set of differential equations (ODE) can be carried out to high precision,
in fact to far more precision than needed to justify the accuracy of the observational constraints on a static model. An alternative method used in a
previous paper 
\citep{AnderSchub:S2007}
can cause precision problems. The method expresses the shape coefficients $f$, $k$ and $h$, or $s _{2i}$,
as polynomials in $\beta$, and forces the polynomial coefficients to satisfy the equations $A_{2 i} = 0$. The problem with this approach
is that a finite number of polynomial coefficients can never be found that satisfy the equations everywhere on the interval $ 0 \leq \beta \leq 1$.
 Numerical compromises must be made in order to satisfy the equations on average over the
 interval. With the ODE approach, the solution for the shape parameters can be automated. 

Using this ODE approach, we first express the shape coefficients as power series in the small rotational parameter $m$.
We illustrate the method for $f$, $k$ and $h$, and use it for the two-zone model, but it can be extended to higher orders as well.
The three spheroidal functions can be written as
\begin{eqnarray}
  f(\beta) &=&   m F_1(\beta) + m^2 F_2(\beta) + m^3 F_3(\beta) \nonumber \\
  k(\beta) &=&   m^2 K_2(\beta) + m^3 K_3(\beta) \nonumber \\
  h(\beta) &=&   m^3 H_3(\beta) 
\label{FKH}
\end{eqnarray}
The first step in the procedure is to substitute these expressions for $f$, $k$ and $h$ into the functions $ \phi_{2 i}$ and $ \phi_{2 i}^\prime$ given in 
Appendix~C and to drop terms of order higher than three. The next step is to substitute the resulting power series into the expressions
for $S _{2 i}$ and $S _{2 i}^\prime$ given by (\ref{Sn}). Finally substitute the resulting gravitational moments and the shape functions $f$,  $k$ and $h$ into the
expressions for $A_2$, $A_4$ and $A_6$ given in Appendix~B.  Then expand to order three in $m$.
This completes the setup of the level-surface problem for the ODE approach.

The lowest-order level-surface potential is $A_2$ to the first order in $m$. Call it $A_{21}$. It is obtained as the
 coefficient of $m$ for $A_2$ from the setup of the problem. It can be written as
\begin{equation}
  A_{21} =   - \frac{1}{2} + S_0 F_1 \left( \beta \right) -\frac{3}{5} \int_\beta^1 \delta \left( z \right) F_1^\prime \left( z \right) \, 
      \mathrm{d}z - \frac{3}{5 \beta^5} \int_0^\beta z^4 \delta \left( z \right) \left( 5 F_1 \left( z \right)
+ z F_1^\prime \left( z \right) \right)\, \mathrm{d}z = 0
\label{A21}
\end{equation}
where $S _0$ can be evaluated by the integral of (\ref{S0}). Differentiate $A _{21}$ once with respect to $\beta$ to obtain
\begin{equation}
  A_{21}^\prime =   \frac{3}{\beta^6} \int_0^\beta z^4 \delta \left( z \right) \left[ 5 F_1 \left( z \right) + z F_1^\prime \left( z \right) \right] \, \mathrm{d} z 
- \frac{1}{\beta} S_0 \left[ 3 F_1 \left( \beta \right) - \beta F_1^\prime \left( \beta \right) \right] = 0
\label{DA21}
\end{equation}
Multiply this derivative by $\beta^6$ and differentiate once again. The result is
\begin{equation}
  \left( \beta^6 A_{21}^\prime \right)^\prime = 6 \beta^4 \delta \left( \beta \right) \left[ F_1 \left( \beta \right)
   + \beta F_1^\prime \left( \beta \right) \right] - \beta^4 S_0 \left[ 6 F_1 \left( \beta \right) - \beta^2 F_1^{\prime \prime} \left( \beta \right) \right] = 0
\label{DDA21}
\end{equation}
This last equation (\ref{DDA21}) is Clairaut's differential equation for the flattening function. However, the first two equations contain integrals that are not known,
and they will enter into higher-order ODE. Therefore, we solve for the two integrals from the two equations 
(\ref{A21}) and (\ref{DA21})
and at the same time solve for the second derivative of  $F_1$ from the third
equation (\ref{DDA21}). This establishes a procedure for all higher orders. The result is
\begin{eqnarray}
 & \int_0^\beta z^4& \delta \left( z \right) \left[ 5 F_1 \left( z \right) + z F_1^\prime \left( z \right) \right]\, \mathrm{d} z =   \beta^5 S_0 F_1 \left( \beta \right) - \frac{1}{3} \beta^6 S_0 F_1^\prime \left( \beta \right) \nonumber \\
 & \int_\beta^1 \delta& \left( z \right) F_1^\prime \left( z \right)\, \mathrm{d} z =   - \frac{5}{6} + \frac{2}{3} S_0 F_1 \left( \beta \right) + \frac{1}{3} \beta S_0 F_1^\prime \left(\beta \right)  \nonumber \\
& F_1^{\prime \prime}& \left( \beta \right) =   \frac{6}{\beta^2} F_1 \left( \beta \right) - \frac{6}{\beta^2} \left( \frac{\delta \left( \beta \right)}{S_0} \right) F_1 \left( \beta \right) - \frac{6}{\beta} \left( \frac{\delta \left( \beta \right)}{S_0} \right) F_1^\prime \left( \beta \right)  
\label{A21Eqs}
\end{eqnarray}
The ODE in (\ref{A21Eqs}) can be solved for $F_1$ and $F_1^\prime$ and the result can be substituted into the two integrals. The solution to the ODE
to first order in $m$ and the corresponding two integrals are now available for higher order ODE. The boundary conditions on the solution are discussed in
section~\ref{SecBC} and they are applied to the two-zone model in section~\ref{2zone}. Note that the density function that completely determines $ F_1$ is given by the ratio $ \delta/S_0$.

The next function for consideration is $K_2$. It is derived from the coefficient $A_{42}$ of  $m^2$ in $A_4$. This time $A_{42}$ is divided by $\beta^2$ and differentiated.
Then the result of that first differentiation is multiplied by $\beta^{10}$ and differentiated once more. This establishes the procedure for all the shape functions.
When $A_6$ is involved, it is divided by $\beta^4$ and differentiated. Then the result of that differentiation is multiplied by $\beta^{14}$ and differentiated once more.
The procedure can in principle be carried to higher orders. For each shape function, three equations are solved for two unknown integrals and the second
derivative of that particular shape function. The sequence of steps for deriving the ODE is $ F_1$, $  K_2$, $  F_2$, $  H_3$, $  K_3$, $  F_3$,
and so forth. The result can be expressed as a nonlinear homogeneous differential equation
plus a function of $\beta$ that is built up by means of the sequence of derivations. We express the ODE in the form
\begin{eqnarray}
  \beta^2 F_i^{\prime \prime} + 6 \beta \left( \frac{\delta}{S_0} \right) F_i^\prime - 6 \left( 1 - \frac{\delta}{S_0} \right) F_i &=&   G_{2 i}~~~~~~i = 1,2,3 \nonumber \\
  \beta^2 K_i^{\prime \prime} + 6 \beta \left( \frac{\delta}{S_0} \right) K_i^\prime - 2 \left( 10 - 3 \frac{\delta}{S_0} \right) K_i &=&   G_{4 i}~~~~~~i = 2,3 \nonumber \\
  \beta^2 H_i^{\prime \prime} + 6 \beta \left( \frac{\delta}{S_0} \right) H_i^\prime - 6 \left( 7 - \frac{\delta}{S_0} \right) H_i &=&   G_{6 i}~~~~~~i = 3
\label{ODE}
\end{eqnarray}
These equations differ in form because of the way $k$ and $h$ are defined in (\ref{r}). The functions $G _{j i}$ are given in Appendix~D.

\subsection{General Boundary Conditions}
\label{SecBC}
The derivatives of the shape functions at the surface where $\beta$ is equal to one can be found sequentially, similar to the technique for finding the ODE.
For the function $ F_1$, the potential $ A_{21}$ is multiplied by $\beta^5$ and differentiated. This is done for $A_{22}$ and $ A_{23}$ as well.
For $ A_{42}$ and $ A_{43}$ the multiplier before differentiation is $\beta^7$, and for $ A_{63}$ it is $\beta^9$. The resulting derivatives are
evaluated for $\beta$ equal to one. Consequently, the integral with limits of integration from $\beta$ to one is set to zero. The integral representing $S _0$
is evaluated at the surface such that $S_0$ is equal to one. The second derivatives are eliminated by substitution of the ODE, again evaluated at the surface.
The resulting first derivatives of the potential functions multiplied by the appropriate $ \beta^i$ can be set to zero and all the derivatives of the shape
functions at the surface boundary can be found sequentially. As a result, the surface boundary conditions  are given by
\begin{eqnarray}
  F_{1 1}^\prime &=&   \frac{5}{2} - 2 F_{1 1} \nonumber \\
  K_{2 1}^\prime &=&   \frac{25}{16} - \frac{5}{4} F_{1 1} -4 K_{21} \nonumber \\
  F_{2 1}^\prime &=&   - \frac{5}{12} + \frac{19}{42} F_{1 1} + \frac{1}{3} F_{11}^2 -2 F_{2 1} + \frac{8}{7} K_{2 1} \nonumber \\
  H_{3 1}^\prime &=&   \frac{25}{8} + \frac{15}{4} F_{1 1} - 5 F_{1 1}^2 - 6 H_{3 1} - 2 K_{2 1} \nonumber \\
  K_{3 1}^\prime &=&   - \frac{25}{24} + \frac{25}{168} F_{1 1} + \frac{137}{168} F_{1 1}^2 - \frac{5}{4} F_{2 1} +\frac{12}{11} H_{3 1} + \frac{904}{231} K_{2 1} - \frac{524}{231} F_{1 1} K_{2 1} - 4 K_{3 1} \nonumber \\
  F_{3 1}^\prime &=&   \frac{155}{72} - \frac{143}{84} F_{1 1} - \frac{47}{147} F_{1 1}^2 + \frac{7}{9} F_{1 1}^3 + \frac{19}{42} F_{21} + \frac{2}{3} F_{1 1} F_{2 1} - 2 F_{3 1} - \frac{92}{77} H_{3 1} \nonumber \\
&&   -\frac{68}{11} K_{2 1} + \frac{22}{2695} F_{1 1} K_{2 1} + \frac{8}{7} K_{3 1}
\label{EqBC}
\end{eqnarray}

This gives the derivatives of the shape functions at the surface. One more set of boundary conditions is needed for a unique solution to the ODE, and hence
for a unique interior model for a given density distribution $\delta(\beta)$, or for a unique EOS that can be integrated by 
(\ref{HEI}) to yield a unique density
distribution. One approach is to iterate on the surface functions, which must satisfy the  boundary conditions given by (\ref{EqBC}), until finite functions
are obtained at the origin. However, this iterative process can be tedious. An alternative, which we adopt here, introduces a constant-density core into the
interior density distribution. The core radius $ \beta_C$ can take on any value in the interval $ 0 < \beta_C \leq 1$, and in principle it can be arbitrarily
small. However, as the core radius approaches zero, the numerical precision required to evaluate the shape functions at the core boundary becomes arbitrarily large.
For every model, there is a practical lower limit to the core radius $ \beta_C$. We illustrate this method of a core plus overlying envelope in Sec.~\ref{2zone}.

\subsection{Calculation of the Normalized Pressure in the Interior}
To the first order in $m$, the pressure depends only on the density distribution. The differential equation for $\chi$ is simply
\citep{ZharkTrubi:PPI1978}
\begin{equation}
  \frac{\mathrm{d}\chi}{\mathrm{d}\beta} =   \left[ - S_0 + \frac{2}{3} m \right] \beta \delta \left( \beta \right) 
\label{dchidbeta}
\end{equation}
The boundary condition for a solution to (\ref{dchidbeta}) is that $\chi$ is equal to zero at $\beta$ equal to one. The density distribution can be piecewise continuous,
as in the two-zone model considered here. However, the pressure and gravitational potential must be continuous over a density discontinuity.
This implies that the spheroidal functions $f$, $k$ and $h$ (or the $s _{2 i}$ functions) and their first derivatives must be continuous throughout the interior. In addition,
the derivative $\mathrm{d}\delta/\mathrm{d}\beta$ must be less than or equal to zero throughout the interior, so that the density either remains constant (incompressible material) or
increases with depth. Also $\delta(\beta)$ must satisfy the boundary condition that the gravitational moment $S _0$ as given by (\ref{S0}) must be equal to one
 at the planet's surface. The surface is defined such that all the planetary mass is contained within the outermost level surface with $\beta$ equal to one.
 Even so, the pressure and the density can be made to match a model atmosphere. The atmosphere is a part of the total mass. In that sense, it is more realistic to define
 the surface at the 100~mbar level in the atmosphere, near the top of the troposphere, not at a more standard one-bar level. Nevertheless, the one-bar level is an acceptable
 approximation to the surface. At least this approximation avoids the complication of treating the atmosphere as a separate zone in the level-surface computation.
 There is something to be said for separating the atmospheric modeling from the interior modeling, and simply making sure the two are consistent at the
 one-bar level. For one thing, the atmosphere is not static, but is dominated by observed zonal flows for all four giant planets in the solar system.
 The theory of figures is a static equilibrium theory. A level surface of one bar in the atmosphere is stretching the static assumption as it is. It is a
 reasonable level to stop the interior modeling. In order that both the density and the pressure go to zero at the surface, the density must go to zero
 at the surface. This introduces another constraint on the interior density distribution. A separate constraint is that the derivative of the density distribution
 at the surface is equal to the derivative in the atmosphere at the one bar level. With the inclusion of the constraint on $ S_0$ previously mentioned,
 this results in a total of three constraints on the interior density distribution. Physically, these three constraints mean that the total mass of the model
 is equal to the measured total mass of the planet, and that the interior density distribution matches the density distribution in the atmospheric model at the one-bar level.

By means of the derivation of the ODE for the shape functions in the interior, it is straightforward to derive the second and third order terms in the
differential equation for the pressure. All the integrals necessary for an evaluation of $ A_0$ are available from the derivation of the ODE. The first order
term in (\ref{dchidbeta}) contains only zero-order shape functions. Similarly, the second order terms in the derivative of $ \beta^2 A_0$ contains
only first order terms in the shape functions. The right side of (\ref{HE}) can be expanded in powers of $m$, and each order can be integrated
separately for purposes of obtaining the normalized pressure $ \chi \left( \beta \right) =   \chi_0 \left( \beta \right) + m \chi_1 \left( \beta \right)
+ m^2 \chi_2 \left( \beta \right) + m^3 \chi_3 \left( \beta \right)$ to third order. The four functions for the integrations are given by
\setlength{\arraycolsep}{.005truein}
\begin{eqnarray}
  \frac{\mathrm{d} \left( \beta^2 A_{00} \right)}{\mathrm{d} \beta} &=&   - \beta S_0 \nonumber \\
  \frac{\mathrm{d}\left( \beta^2 A_{01} \right)}{\mathrm{d} \beta} &=&   \frac{2}{3} \beta \nonumber \\
  \frac{\mathrm{d} \left( \beta^2 A_{02} \right)}{\mathrm{d}\beta} &=&   \frac{8}{45} \beta \left( 2 F_1 + \beta F_1^\prime \right) + \frac{4}{45} \beta S_0 \left( 5 F_1^2 + 2 \beta F_1 F_1^\prime + \beta^2 F_1^{\prime 2} \right) \nonumber \\
  \frac{\mathrm{d}\left( \beta^2 A_{03} \right)}{\mathrm{d}\beta} &=&   - \frac{8}{135} \beta \left[ 5 F_1^2 - 6 F_2 + 2 \beta F_1 F_1^\prime + \beta \left( \beta F_1^{\prime 2} - 3 F_2^\prime \right) \right] \nonumber \\
&&   + \frac{4}{2835} \beta S_0 \left( 385 F_1^3 + 231 \beta F_1^2 F_1^\prime \right) \nonumber \\
&& + \frac{24}{2835} \beta^2 S_0 F_1^\prime \left[ 21 F_2 + 12 K_2 + \beta \left( 2 \beta F_1^{\prime 2} + 21 F_2^\prime + 12 K_2^\prime \right)  \right] \nonumber \\
&& + \frac{24}{2835} \beta S_0 F_1 \left[ 105 F_2 + 60 K_2 + \beta \left( 25 \beta F_1^{\prime 2}+ 21 F_2^\prime + 12 K_2^\prime \right) \right] 
\label{Dbet2A0}
\end{eqnarray}
The pressure can be found by multiplying the four derivatives in (\ref{Dbet2A0}) by the normalized density
$\delta \left( \beta \right)$ and integrating, with the boundary condition $\chi \left(1 \right)$ equal to zero. 

The method described here can be applied to the simple case of a planet made up of incompressible material in HE.
The normalized density is a constant equal to one, and the zero degree gravitational moment $S_0$ is also a constant equal to one.
The ODE simplify considerably, but that fact can be ignored, and our general
numerical procedure can be applied to the constant-density case.
As a result, the numerical solution to the ODE yields the result 
\begin{eqnarray}
  f &=&   \frac{5}{4} m \left( 1 + \frac{15}{56} m + \frac{925}{1568} m^2 \right) \nonumber \\
  k &=&   0 \nonumber \\
  h &=&   0
\label{fkhConst}
\end{eqnarray}    
The normalized axial moment of inertia $C/Ma^2$ for this configuration is equal to 2/5, independent of $m$.
A numerical integration of the four pressure functions in (\ref{Dbet2A0}), with $ \delta \left( \beta \right)$ equal to one,
yields the following result for the normalized pressure
\begin{equation}
  \chi =   \left( 1 - \beta^2 \right) \left( \frac{1}{2} - \frac{1}{3} m - \frac{41}{72} m^2 - \frac{1235}{2268} m^3 \right)
\label{chiConst}
\end{equation}  
In the above, the real numbers returned by the numerical integration have been replaced by nearby rational numbers with small denominator.
This has only been done in (\ref{chiConst}).

The next simplest density distribution is the linear distribution. Because it implies compressible material, it is a far better approximation to a real planet
than the constant-density model. The normalized density $\delta \left( \beta \right)$ is equal to $ 4 \left( 1 - \beta \right)$. For purposes of applying our
numerical procedure, we introduce a core of normalized radius $ \beta_C$ equal to 0.05. The normalized constant density in the core is equal to 3.85.
The gravitational moment $S_0$ in the envelope is equal to $4 - 3 \beta$. In this model, there is a negligible fractional core mass equal to
77/160000. The numerical integration is carried out in the envelope over the interval $0.05 \leq \beta \leq 1$.     

\subsection{Calculation of the Coefficients $ J_n$ in the Exterior Gravitational Potential}
\label{J2J4J6}
The solution to the differential equations to third order in the small rotational parameter $m$ yields the six shape functions $ F_1$,
$F_2$, $F_3$,  $K_2$, $K_3$, $H_3$.
If good observations of the shape of the planet are available, such as for the Earth, these shape functions can be used directly to constrain the envelope
density $\delta_E$. However, for the outer planets, the measured zonal gravitational coefficients $ J_2$, $ J_4$ and $ J_6$
provide far better constraints on $\delta_E$. The calculated values of the three coefficients are given by (\ref{Jn}). 

The procedure for finding values of the gravitational coefficients in terms of the shape functions
is to first express the coefficients as a truncated power series in $m$ according to
\begin{eqnarray}
  J_2 &=&   m J_{2 1} + m^2 J_{2 2} + m^3 J_{23} \nonumber \\
  J_4 &=&   m^2 J_{4 2} + m^3 J_{43} \nonumber \\
  J_6 &=&   m^3 J_{6 3}
\label{Jij}
\end{eqnarray}
Next we recognize that the coefficients, when referenced to the equatorial radius $a$, are proportional to the gravitational moments $S_n$ by
\begin{equation}
  S_n =   \left( \frac{a}{s} \right)^n J_n
\label{SnvsJn}
\end{equation}
and where, from (\ref{rmu}) with $\mu$ set equal to zero
\begin{equation}
  \frac{a}{s} =   1 + \frac{1}{3} f + \frac{2}{9} f^2 + \frac{14}{81} f^3 + \frac{26}{105} h + \frac{8}{15} k + \frac{16}{63} f k
\label{abys}
\end{equation}
Substitute (\ref{SnvsJn}) into the expressions for the potential functions $A_2$,$ A_4$,$ A_6$  given respectively by
(\ref{A2}), (\ref{A4}), (\ref{A6}), and evaluate at the surface. Use the truncated series of (\ref{Jij}) for the $ J_n$ and the similar series for the
shape functions given in (\ref{FKH}). The functions $S_n^\prime$ are all zero at the surface. Expand to order three in $m$. Since all the
coefficients of powers of $m$ are zero, this process yields six equations which can be solved for the six $ J_n$ functions in terms
of the six shape functions from the differential equations, again evaluated at the surface. The result is
\begin{eqnarray}
  J_{21} &=&   -\frac{1}{3} + \frac{2}{3} F_{11} \nonumber \\
  J_{22} &=&   \frac{2}{21} F_{11} - \frac{1}{3} F_{11}^2 + \frac{2}{3} F_{21} + \frac{8}{21} K_{21} \nonumber \\
  J_{23} &=&   - \frac{11}{147} F_{11}^2 + \frac{2}{21} F_{21} - \frac{2}{3} F_{11} F_{21} + \frac{2}{3} F_{31} - \frac{2}{21} H_{31} - \frac{16}{105} K_{21} + \frac{40}{147} F_{11}K_{21} + \frac{8}{21} K_{31} \nonumber \\
  J_{42} &=&   \frac{4}{7} F_{11} - \frac{4}{5} F_{11}^2 - \frac{32}{35} K_{21} \nonumber \\
  J_{43} &=&   - \frac{22}{49} F_{11}^2 + \frac{4}{5} F_{11}^3 + \frac{4}{7} F_{21} - \frac{8}{5} F_{11} F_{21} - \frac{192}{385} H_{31} + \frac{208}{385} K_{21} + \frac{3616}{2695} F_{11} K_{21} - \frac{32}{35} K_{31} \nonumber \\
  J_{63} &=&   - \frac{20}{21} F_{11}^2 + \frac{8}{7} F_{11}^3 + \frac{80}{231} H_{31} - \frac{160}{231} K_{21} + \frac{128}{77} F_{11} K_{21}
\label{JvsFKH}
\end{eqnarray}

\subsection{The Two-Layer Model}
\label{2zone}
The two-layer model consists of a constant density core of normalized density $\delta_C$, plus an envelope of normalized density $ \delta_E$.
The envelope density can be a function of $\beta$, or even piecewise continuous in two or more zones overlying the constant-density core.
The two densities are connected by means of the mass constraint implied by (\ref{S0}), and they must satisfy the following equation
\begin{equation}
  \delta_C \beta_C^3 + 3 \int_{\beta_C}^1 \beta^2 \delta_E \left( \beta \right)\,\mathrm{d} \beta = 1
\label{MC}
\end{equation}
A particular interior model is defined by the envelope density $ \delta_E$ and the core radius $ \beta_C$. The core density $ \delta_C$ is a
derived constant obtained from (\ref{MC}). As the core radius approaches zero, the core density approaches positive infinity. However, the core
mass given by $ \delta_C \beta_C^3$ is finite at the origin. For $ \beta_C$ arbitrarily small, the core mass can represent a point-mass core with
mass greater than or equal to zero. Whatever the values for $ \delta_E ( \beta )$ and $  \beta_C$, the ODE of (\ref{ODE}) can be solved exactly
in the core, and the second set of boundary conditions for the envelope integration can be found at the core-envelope boundary.

\subsection{Solution to the Theory of Figures in a Constant-Density Core}
\label{coresol}
The functions needed for the ODE of (\ref{ODE}) are $S_0$ and $ \delta/S_0$. For a constant-density core, $ S_0$ is
simply $ \delta_C$ and the ratio $ \delta/S_0$ is one. It follows from (\ref{ODE})  and 
(\ref{Gji}) that the first-order flattening function
$F_1$ is a constant. It has the value $ F_{1B}$ everywhere in the core and its derivative is zero within the core. This simplifies
the other ODE of (\ref{ODE}) considerably. The equation for $ K_2$ is
\begin{equation}
  \beta^2 K_2^{\prime \prime} + 6 \beta K_2^\prime - 14 K_2 = 0
\label{K2ODE}
\end{equation}
The boundary conditions on (\ref{K2ODE}) are that $ K_2$ is finite at the origin and that it is equal to $ K_{2B}$ on the core-envelope
boundary. The solution to (\ref{K2ODE})  and the boundary condition at $\beta$ equal to $ \beta_C$ are
\begin{eqnarray}
  K_2 &=&   K_{2 B} \left( \frac{\beta}{\beta_C} \right)^2 \nonumber \\
  K_{2B}^\prime &=& 2 \left( \frac{K_{2 B}}{\beta_C} \right)  
\label{K2Sol}
\end{eqnarray}
Both the shape functions and their first derivatives are continuous at the core-envelope boundary.
Therefore, the above boundary condition applies to both the core and the envelope at $\beta$ equal to $  \beta_C$.
Similarly, the equation for $F_2$ from (\ref{ODE}) and (\ref{Gji}) is
\begin{equation}
  \beta^2 F_2^{\prime \prime} + 6 \beta F_2^\prime  =   - 8 K_2
\label{F2ODE}
\end{equation}
with the solution 
\begin{eqnarray}
  F_2 &=&   F_{2 B} + \frac{4}{7} K_{2 B} \left[1 - \left(  \frac{\beta}{\beta_C} \right)^2 \right] \nonumber \\
  F_{2B}^\prime &=& - \frac{8}{7} \left( \frac{K_{2 B}}{\beta_C} \right)  
\label{K2Solsec}
\end{eqnarray}
The equation for $H_3$ is
\begin{equation}
  \beta^2 H_3^{\prime \prime} + 6 \beta H_3^\prime  - 36 H_3 =   - 88 \left( \frac{\beta}{\beta_C} \right)^2 F_{1B} K_{2B} 
\end{equation}
with solution
\begin{eqnarray}
  H_3 &=&   H_{3 B} \left( \frac{\beta}{\beta_C} \right)^4 + 4 F_{1 B} K_{2 B} \left( \frac{\beta}{\beta_C} \right)^2  \left[ 1 - \left( \frac{\beta}{\beta_C} \right)^2 \right]  \nonumber \\
  H_{3B}^\prime &=& 4 \left( \frac{H_{3 B}}{\beta_C} \right) - 8 \left( \frac{F_{1 B} K_{2 B}}{\beta_C} \right) 
\label{H3Sol}
\end{eqnarray}
The equation for $K_3$ is
\begin{equation}
  \beta^2 K_3^{\prime \prime} + 6 \beta K_3^\prime  - 14 K_3 =   48 \left( \frac{\beta}{\beta_C} \right)^4 F_{1B} K_{2B} - 12 \left( \frac{\beta}{\beta_C} \right)^4 H_{3 B}
\end{equation}
with solution
\begin{eqnarray}
  K_3 &=&   K_{3 B} \left( \frac{\beta}{\beta_C} \right)^2 - \frac{24}{11} F_{1 B} K_{2 B} \left( \frac{\beta}{\beta_C} \right)^2  \left[ 1 - \left( \frac{\beta}{\beta_C} \right)^2 \right] 
+ \frac{6}{11} H_{3 B} \left( \frac{\beta}{\beta_C} \right)^2  \left[ 1 - \left( \frac{\beta}{\beta_C} \right)^2 \right]\nonumber \\
  K_{3B}^\prime &=& 2 \left( \frac{K_{3 B}}{\beta_C} \right) + \frac{48}{11} \left( \frac{F_{1 B} K_{2 B}}{\beta_C} \right) - \frac{12}{11} \left( \frac{H_{3 B}}{\beta_C} \right) 
\label{K3Sol}
\end{eqnarray}
Finally, the equation for $F_3$ is
\begin{eqnarray}
  \beta^2 F_3^{\prime \prime} + 6 \beta F_3^\prime  &=&   \frac{8}{385} \left( \frac{\beta}{\beta_C} \right)^2 \left[ 2039 - 3150 \left( \frac{\beta}{\beta_C} \right)^2  \right] F_{1B} K_{2B} \nonumber \\
&& - \frac{12}{11} \left( \frac{\beta}{\beta_C} \right)^2 \left[ 4 - 15 \left( \frac{\beta}{\beta_C} \right)^2  \right] H_{3 B} - 8 \left( \frac{\beta}{\beta_C} \right)^2 K_{3 B} 
\end{eqnarray}
with solution
\begin{eqnarray}
  F_3 &=&   F_{3 B} - F_{1 B} K_{2 B} \left[ 1 -  \left( \frac{\beta}{\beta_C} \right)^2  \right] \left[ \frac{296}{245} - \frac{20}{11} \left( \frac{\beta}{\beta_C} \right)^2 \right] \nonumber \\
&& - H_{3 B} \left[ 1 -  \left( \frac{\beta}{\beta_C} \right)^2  \right] \left[ \frac{1}{7} + \frac{5}{11} \left( \frac{\beta}{\beta_C} \right)^2 \right] +\frac{4}{7} K_{3 B} \left[ 1 -  \left( \frac{\beta}{\beta_C} \right)^2  \right] \nonumber \\
  F_{3B}^\prime &=&   - \frac{3288}{2695} \left( \frac{F_{1 B} K_{2 B}}{\beta_C} \right) + \frac{92}{77} \left( \frac{H_{3 B}}{\beta_C} \right) - \frac{8}{7} \left( \frac{K_{3 B}}{\beta_C} \right)
\label{F3Sol}
\end{eqnarray}

\subsection{Solution to the Theory of Figures in a Constant-Density Envelope}
\label{envsol}
When the normalized envelope density $ \delta_E$ is a constant, (\ref{MC}) yields the following expression for the normalized density in the core
\begin{equation}
  \delta_C =   \delta_E +\frac{1 - \delta_E}{\beta_C^3}
\label{delC}
\end{equation}
The gravitational moment $S_0$ in the envelope is similarly
\begin{equation}
  S_0 =   \delta_E +\frac{1 - \delta_E}{\beta^3}
\label{S0C}
\end{equation}
A value for $  \delta_E$, and also $ S_0$ from (\ref{S0C}), can be substituted into the differential equations
given by (\ref{ODE})  and (\ref{Gji}),
and the solution for the spheroidal functions in the envelope can be found subject to the boundary conditions at the surface, given in Sec.~\ref{SecBC},
and the boundary conditions at normalized radius $  \beta_C$ given by expressions at the core boundary in Sec.~\ref{coresol}. The normalized core radius
$  \beta_C$ enters through the boundary conditions of Sec.~\ref{coresol} only. For any finite value of $  \delta_E$ the solutions to the ODE are complicated
and lengthy, although solutions do exist for the two-zone model considered here. Nevertheless, beyond this idealized two-zone model, numerical integration is
required when the envelope density varies with $\beta$. Perhaps the exact solutions to the ODE for constant density are useful for purposes of checking the
precision of the numerical integration, but they have no particular advantage to the problem of interior modeling. In practice, numerical integration is a useful
general approach for any envelope density, including constant density. The precision of the numerical integration can be adjusted such
that it is competitive with the exact solutions to the constant-density envelope, especially given the limited accuracy required by the observational constraints.

However, the exact solution is tractable for the case where the density in the envelope is zero. This is the generalized Roche model considered by
\citet{ZharkTrubi:PPI1978} and discussed in Section~\ref {sec:comp}.
The envelope contains no gravitational mass contribution to the HE, but there is an inertial contribution from the centrifugal potential,
 and hence a contribution to the surface shape of the planet. For purposes of illustrating the ODE approach, we solve this Roche case
 in section~\ref{Roche} and finally consider the general case of finite density in section~\ref{ConstCase}.

\subsubsection{The Generalized Roche Model}
\label{Roche}
For $  \delta_E$ equal to zero, the density of material in the core is given by $  \rho_0/\beta_C^3$. In turn, the differential equation for 
$F_1$ in the envelope becomes
\begin{equation}
  \beta^2 F_1^{\prime \prime} - 6 F_1 =   0
\label{F1Roche}
\end{equation}
The boundary conditions on the solution for $F  _1$ is that $ F_1^\prime = 5/2 - 2 F_1$ at $\beta$ equal to one,
and $ F_1^\prime$ equal to zero at $\beta$ equal to $ \beta_C$. With these boundary conditions,  $F  _1$ in the envelope is given by
\begin{eqnarray}
  F_1 &=& \frac{2 \beta^5 + 3 \beta_C^5}{4 \beta^2} \nonumber \\
  F_1^\prime &=& \frac{3 \left( \beta^5 - \beta_C^5 \right)}{2 \beta^3} \nonumber \\
  F_{1 1} &=&   \frac{1}{4} \left( 2 + 3 \beta_C^5 \right) \nonumber \\
  F_{1 B} &=&   \frac{5}{4} \beta_C^3  
\label{SolF1Roche}
\end{eqnarray}
Here $ F_{1 1}$ is the value of $  F_1$ at the surface, and $  F_{1 B}$ is the value of $  F_1$ at the core boundary.
All the quantities in (\ref{SolF1Roche}) are needed for solutions to the ODE for the higher order shape functions.
The value of $  F_1 $ in the core is determined by the envelope density distribution, which in this Roche case is zero,
but it is true in general for any envelope density distribution. For the Roche model, the density distribution
in the core is simply a constant, given by $F_{1 B}$ according to the core solution of Sec.~\ref{coresol}.

The differential equation for the function $ K_2$ is obtained from (\ref{ODE}) and (\ref{Gji}),
and after the envelope density is set to zero and the solution for $ F_1$ is inserted into $ G_{4 2}$,
the equation for the Roche model is
\begin{equation}
  \beta^2 K_2^{\prime \prime} - 20 K_2 =   \frac{15}{16} \beta \left( \beta^5 + 4 \beta_C^5 \right)
\label{F1Rochesec}
\end{equation}
Again, with the boundary conditions from Sections~\ref{SecBC} and \ref{coresol}, the solution for $K_2$ is
\begin{eqnarray}
  K_2 &=&   \frac{3}{32} \frac{\left( \beta^5 - \beta_C^5\right)^2}{\beta^4} \nonumber \\
  K_2^\prime &=&   \frac{3}{16} \frac{\left( 3 \beta^{10} - \beta^5 \beta_C^5 - 2 \beta_C^{10} \right)}{\beta^5} \nonumber \\
  K_{2 1} &=&   \frac{3}{32} \left( 1 - \beta_C^5 \right)^2 \nonumber \\
  K_{2 B} &=&   0  
\label{SolK2Roche}
\end{eqnarray}
Similarly, the differential equation for $F_2$ is
\begin{equation}
  \beta^2 F_2^{\prime \prime} - 6 F_2 =   - \frac{3 \beta_C^5}{16 \beta^4} \left( 4 \beta^5 + 11 \beta_C^5 \right)
\label{F2Roche}
\end{equation}
and the solution is
\begin{eqnarray}
  F_2 &=&    \frac{\beta_C^5}{224 \beta^4} \left( 28 \beta^5 + 80 \beta^2 \beta_C^3 - 33 \beta_C^5 \right)    \nonumber \\
  F_2^\prime &=&   \frac{\beta_C^5}{56 \beta^5} \left( 7 \beta^5 - 40 \beta^2 \beta_C^3 + 33 \beta_C^5 \right)   \nonumber \\
  F_{2 1} &=&   \frac{\beta_C^5}{224} \left( 28 + 80 \beta_C^3 -33 \beta_C^5 \right)  \nonumber \\
  F_{2 B} &=& \frac{75 \beta_C^6}{224}      
\label{SolF2Roche}
\end{eqnarray}
This process can be extended to third order, although the expressions for $ H_3$, $ K_3$ and $ F_3$
as a function of $\beta$ in the envelope become more lengthy. We list here only their values at the surface, which are
\setlength{\arraycolsep}{.0025truein}
\begin{eqnarray}
\!\! \!\!\! H_{31} &=&   \frac{1}{160}\!  \left( 1\! -\! \beta_C^5 \right)^2 \left( 38\! +\! 67 \beta_C^5 \right) \nonumber \\
\!\! \!\!\! K_{31} &=&   - \frac{1}{2240}\! \left( 42\! +\! 49 \beta_C^5\! +\! 200 \beta_C^8\! -\! 424 \beta_C^{10} - 200 \beta_C^{13}\! +\! 333 \beta_C^{15} \right) \nonumber \\
\!\!\! \!\!\! F_{31} &=&   \frac{1}{94080}\! \left( 7056\! +\! 392 \beta_C^5 \!+\! 5600 \beta_C^8\! -\! 462 \beta_C^{10}\! +\! 74000 \beta_C^{11}\! -\! 13200 \beta_C^{13}\! -\! 4011 \beta_C^{15} \right)  \quad
\label{H31K31F31Roche}
\end{eqnarray}

The values for the gravitational coefficients can be obtained from (\ref{JvsFKH}). The results are
\begin{eqnarray}
  J_{21} &=&   \frac{1}{2} \beta_C^5 \nonumber \\
  J_{22} &=&   - \frac{1}{2} \beta_C^5 \left( \frac{1}{3} - \frac{10}{21} \beta_C^3 + \frac{1}{2} \beta_C^5 \right) \nonumber \\
  J_{23} &=&   - \frac{1}{2} \beta_C^5 \left( \frac{23}{180} + \frac{10}{63} \beta_C^3 - \frac{1}{30} \beta_C^5 -\frac{925}{882} \beta_C^6 + \frac{10}{21} \beta_C^8 + \frac{9}{140} \beta_C^{10} \right) \nonumber \\
  J_{42} &=&   - \frac{15 \beta_C^{10}}{28} \nonumber \\
  J_{43} &=&   \frac{15 \beta_C^{10}}{28} \left( \frac{2}{3} - \frac{20}{21} \beta_C^3 + \beta_C^5 \right) \nonumber \\
  J_{63} &=& \frac{125}{168} \beta_C^{15} 
\label{Js4Roche}
\end{eqnarray}
These results for the gravitational coefficients in the generalized Roche model agree with
\citet{ZharkTrubi:PPI1978},
except for $ J_{23}$. Total agreement is a good check on our ODE method, as the derivation in
\citet{ZharkTrubi:PPI1978}
 is quite different from ours.
We suggest that the third order term for $ J_2$ in 
\citet{ZharkTrubi:PPI1978}
contains typographical errors.
For example, by setting the core radius to one in (\ref{Js4Roche}),
the case of a constant-density planet is recovered. The result is
\begin{eqnarray}
  J_2 &=&   \frac{1}{2} m - \frac{5}{28} m^2 + \frac{25}{196} m^3 \nonumber \\
  J_4 &=&   - \frac{15}{28} m^2 + \frac{75}{196} m^3 \nonumber \\
  J_6 &=&   \frac{125}{168} m^3
\label{Js4Const}
\end{eqnarray}
This is correct 
\citep{ZharkTrubi:PPI1978},
 and it is a good check on the ODE method. However, if $ \beta_C$ is set equal to one in
(34.6) for $ J_2$ in 
\citet{ZharkTrubi:PPI1978},
 the result is $ J_2 = (1/2) m - (5/28) m^2 - (1889/4410) m^3 $.
This is not correct. We conclude that there is agreement between our ODE method and the method of
Zharkov and Trubitsyn, but only if the third order $ J_2$ term in 
\citet{ZharkTrubi:PPI1978} is brought into agreement
with $ J_{23}$ in (\ref{Js4Roche}).

Although the generalized Roche model is an idealization of a real giant planet, it illustrates the method.
Starting with a density distribution in the envelope given by $ \delta_E ( \beta )$ and a core radius $ \beta_C$,
the zonal gravitational coefficients in the external gravitational potential can be calculated. In general, the results are
obtained by numerical integration of the ODE, but the numerical values analogous to the six functions of (\ref{Js4Roche})
can be calculated to any arbitrary precision. A comparison of the calculated values with the measured values is achieved
by calculating the value of the small rotational parameter $m$ for the planet in question, and then by applying (\ref{Jij}).

\subsubsection{Model for a Finite-Density Envelope}
\label{ConstCase}
\begin{figure}
\includegraphics[scale=1.2]{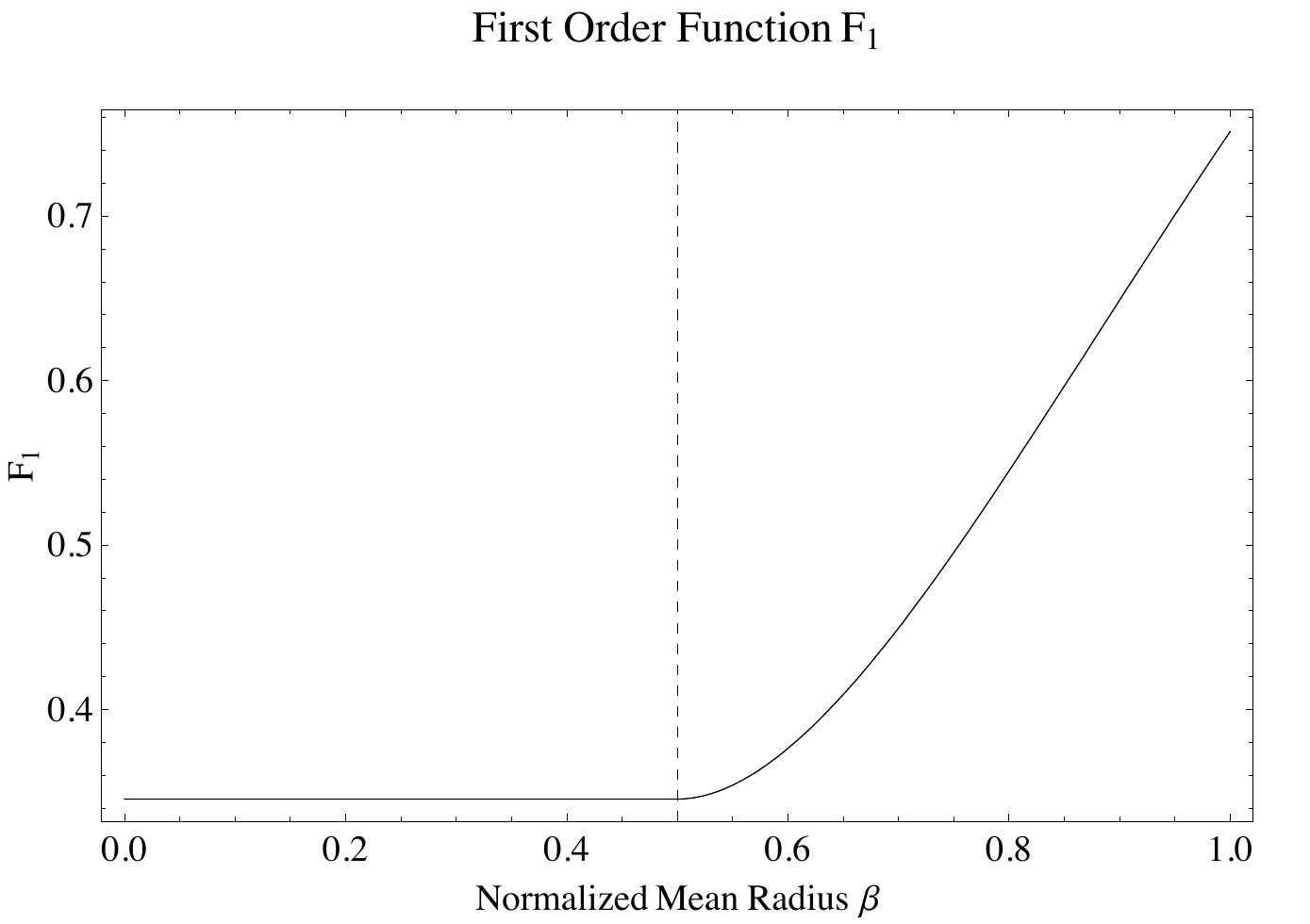}
\caption{Function  $F_1 (\beta)$ for normalized envelope density $\delta_E$ of 0.5 and core radius $\beta_C$ of 0.5.}
\label{fig:3}       
\end{figure}

Even for this simple case of a finite-density envelope, the solution to the ODE can be obtained by numerical integration.
 For purposes of illustrating the method, we pick a normalized envelope density of 1/2 and a normalized core radius of 1/2.
 By (\ref{MC}) the density $ \delta_C$ in the core is equal to 9/2, and the percentage of the total mass in the core
 ($ \delta_C \beta_C^3$) is 9/16. This particular choice of $ \delta_E$ and $ \beta_C$
 results in a fairly simple differential equation for the first-order function $ F_1$. By (\ref{ODE})
 we have
\begin{equation}
  \beta^2 \left( 1 + \beta^3 \right) F_1^{\prime \prime} + 6 \beta^4 F_1^\prime - 6 F_1 =   0
\label{F1ODESimp}
\end{equation}
The integration can be done numerically subject to the boundary conditions of sections~\ref{SecBC} and \ref{coresol}, which for $ F_1(\beta)$ are
\begin{eqnarray}
  F_1^\prime \left( 1 \right) &=&   \frac{5}{2} - 2 F_1 \left( 1 \right) \nonumber \\
  F_1^\prime \left( \beta_C \right) &=&   0
\label{F1BC}
\end{eqnarray}
The limits of integration are from $ \beta_C$ to one. After the numerical integration is complete,
a value of $ F_1$ anywhere on the interval $ \beta_C \leq \beta \leq 1$ can be found by numerical interpolation.
This solution in the envelope can be matched to the solution in the core given in Sec.~\ref{coresol}.
A plot of this particular case throughout the interior is shown in Fig.~\ref{fig:3}.

\begin{figure}
\includegraphics[scale=1.2]{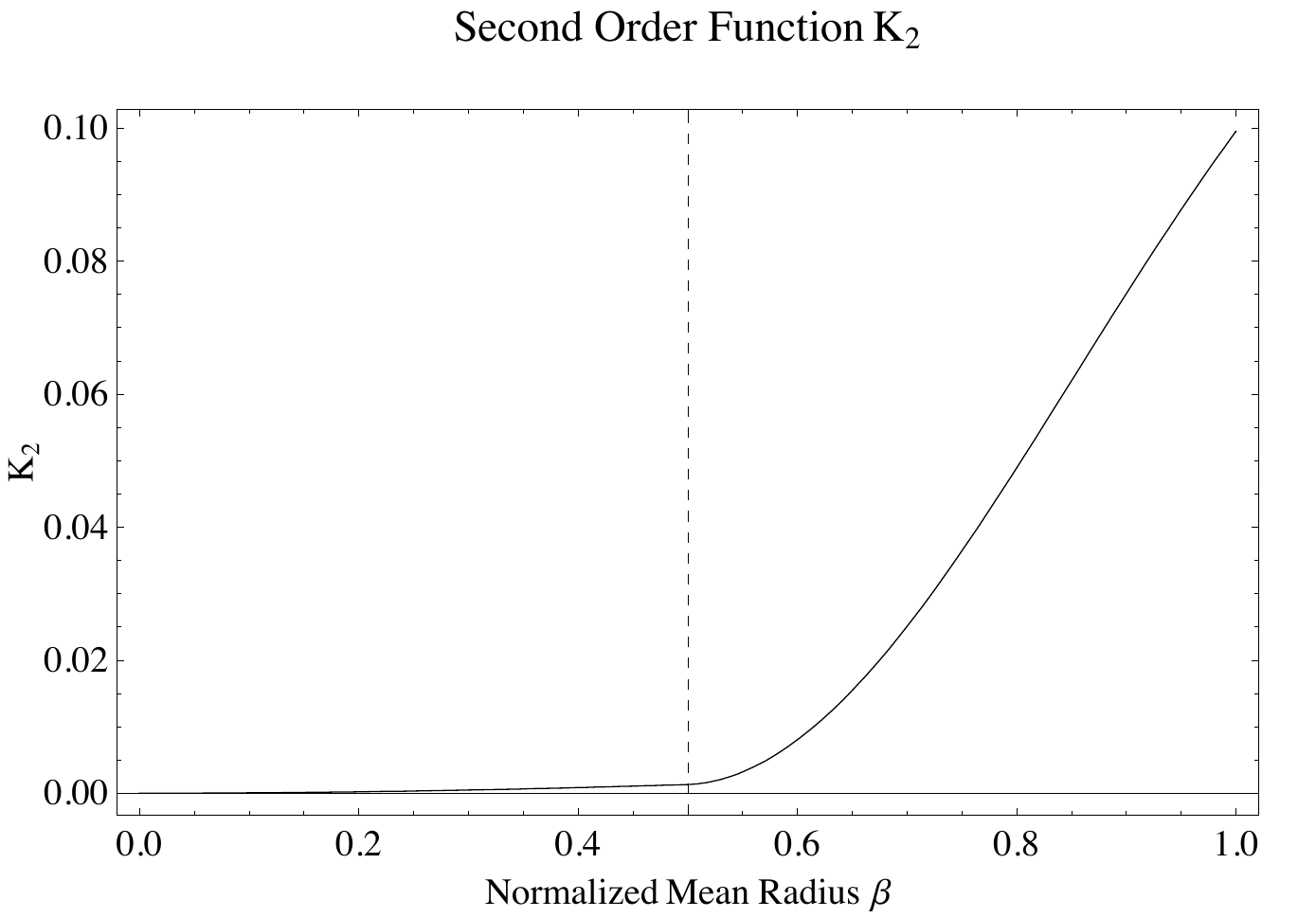}
\caption{Function $K_2 (\beta)$ for normalized envelope density $\delta_E$ of 0.5 and core radius $\beta_C$ of 0.5. }
\label{fig:4}       
\end{figure}

Because the differential equation for $ K_2$ involves both $ F_1$ and its first derivative, it must be evaluated
numerically by interpolating in the numerical solution to (\ref{F1ODESimp}). Furthermore, the boundary condition
at the surface is not known until $ F_1$ at the surface is known. Therefore, we do not write down the differential equation that must be
integrated, but instead numerically evaluate it according to (\ref{ODE}) on the interval $ \beta_C \leq \beta \leq 1$.
The boundary conditions for this special case, with $ \delta_E$ and $  \beta_C$ both equal to 1/2, are obtained from the
expressions given in Sec.~\ref{SecBC} and Sec.~\ref{coresol}, and include the solution for $ F_1$ at the surface.
In general, the boundary conditions are
\begin{eqnarray}
  K_2^\prime \left( \beta_C \right) &=&   2 \left( \frac{K_2 \left( \beta_C \right)}{\beta_C} \right) \nonumber \\
  K_2^\prime \left( 1 \right) &=&   \frac{25}{16} - \frac{5}{4} F_1 \left( 1 \right) - 4 K_2 \left( 1 \right)
\label{K2BC}
\end{eqnarray}
and by numerical interpolation in the previous solution, $ F_1 \left( 1 \right)$ is equal to 99060576/131853043,
accurate to 16 places past the decimal. With these boundary conditions, numerical integration yields the solution for
$ K_2$ in the envelope, which can be matched to the core solution and plotted. The result is shown in Fig.~\ref{fig:4}.

\begin{figure}
\includegraphics[scale=1.2]{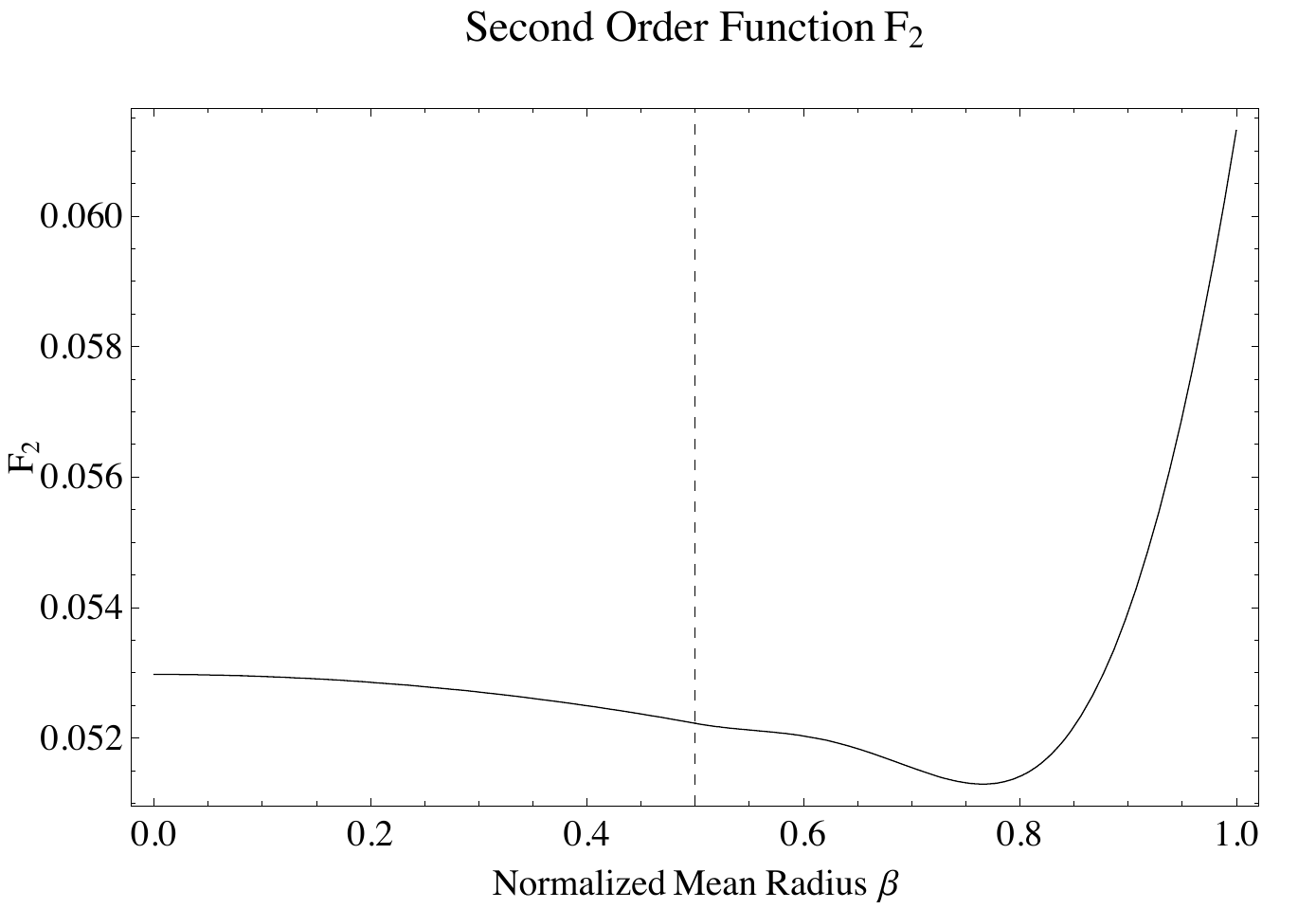}
\caption{Function $F_2 (\beta)$ for normalized envelope density $\delta_E$ of 0.5 and core radius $\beta_C$ of 0.5.}
\label{fig:5}       
\end{figure}

The procedure is similar for the function $ F_2$, and the differential equation from (\ref{ODE})
involves the previous solution for both $F_1$ and $ K_2$ and their first derivatives. The boundary conditions are
\begin{eqnarray}
  F_2^\prime \left( \beta_C \right) &=&   -\frac{8}{7} \left( \frac{K_2 \left( \beta_C \right)}{\beta_C} \right) \nonumber \\
  F_2^\prime \left( 1 \right) &=&   - \frac{5}{12} + \frac{19}{42} F_1 \left( 1 \right) + \frac{1}{3} F_1 \left( 1 \right)^2
  + \frac{8}{7} K_2 \left( 1 \right) - 2 F_2 \left( 1 \right)  
\label{F2BC}
\end{eqnarray}
where for this special case, $K_2 \left( 1/2 \right)$ is equal to 143636/109713139
and $K_2 \left( 1 \right)$ is equal to 6168175/61992373, again accurate to 16 places
past the decimal. After numerical integration, the solution for $F_2$ is represented by Fig.~\ref{fig:5}. 

\begin{figure}
 \includegraphics[scale=1.2]{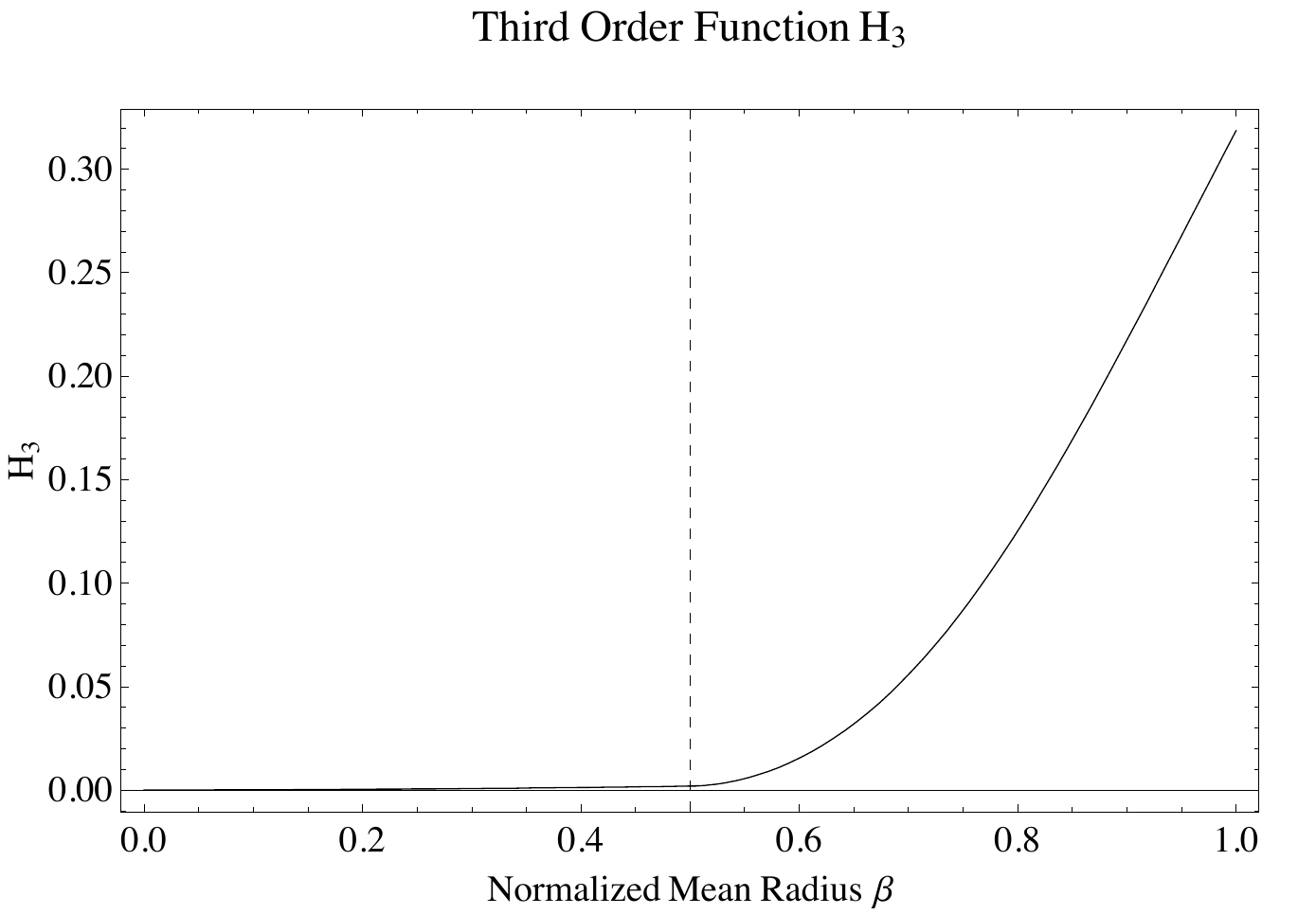}
\caption{Function $ H_3 (\beta)$ for normalized envelope density $\delta_E$ of 0.5 and core radius $\beta_C$ of 0.5. }
\label{fig:6}       
\end{figure}

\begin{figure}
  \includegraphics[scale=1.2]{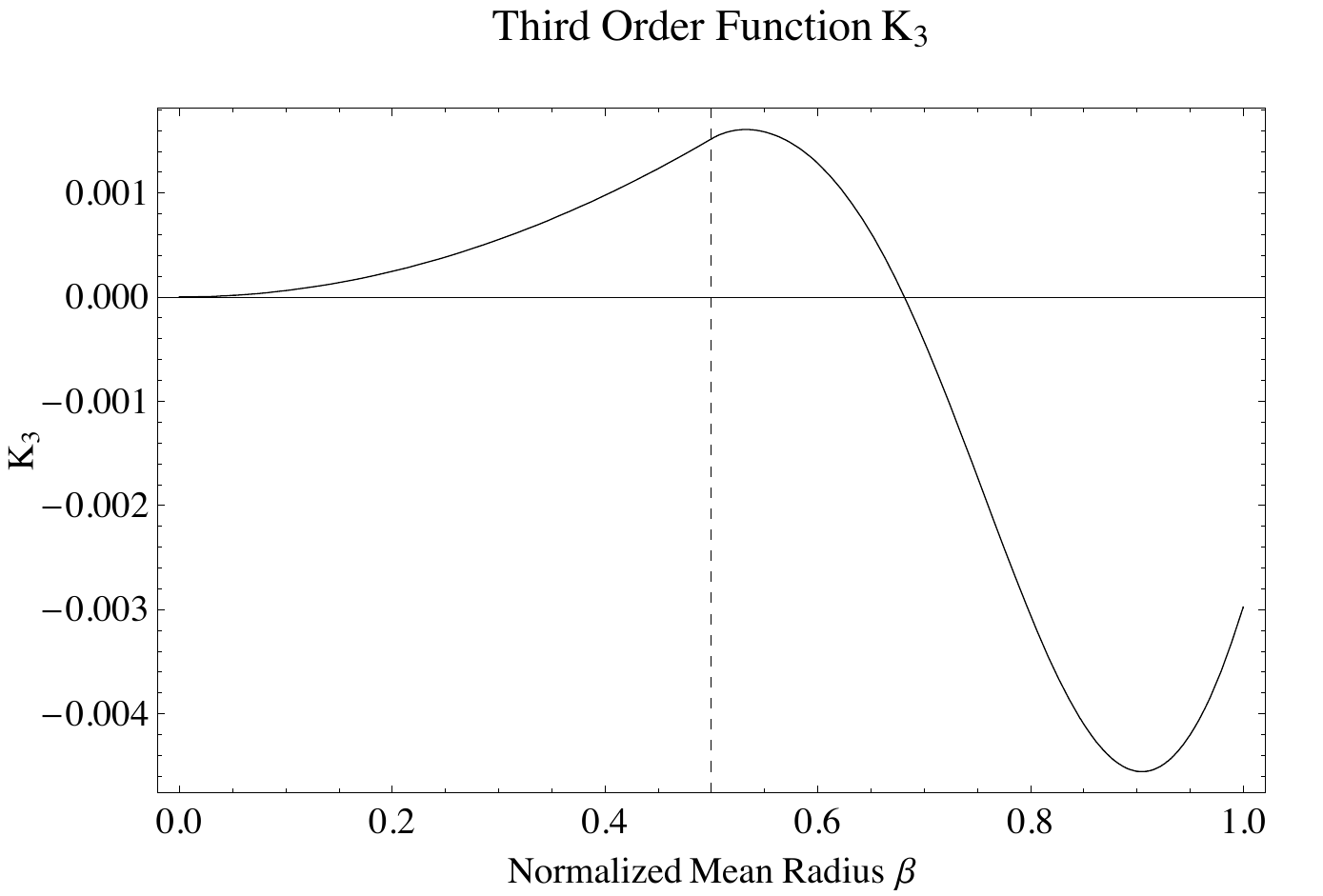}
\caption{Function $K_3 (\beta)$ for normalized envelope density $\delta_E$ of 0.5 and core radius $\beta_C$ of 0.5.}
\label{fig:7}       
\end{figure}

\begin{figure}
  \includegraphics[scale=1.2]{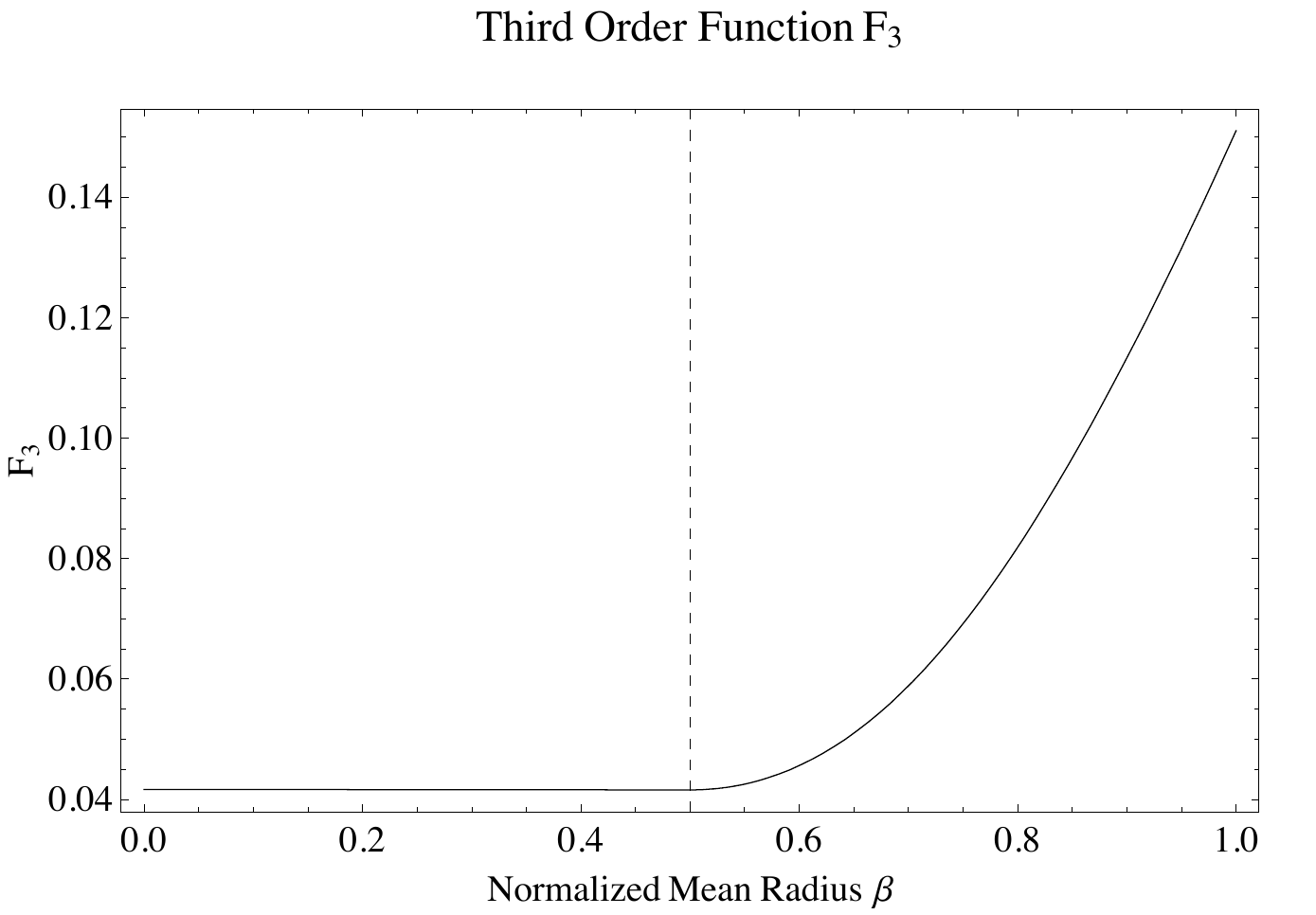}
\caption{Function $F_3 (\beta)$ for normalized envelope density $\delta_E$ of 0.5 and core radius $\beta_C$ of 0.5. }
\label{fig:8}       
\end{figure}

The above process can be repeated for the third-order functions $H_3$, $ K_3$ and $ F_3$,
 in that order. As each function is introduced, all previous solutions are used in both the ODE and in the
 boundary conditions. The results for the special case considered here are represented by
 Fig.~\ref{fig:6}, Fig.~\ref{fig:7} and Fig.~\ref{fig:8}. All six plotted functions can be evaluated at the surface.
 As a result, the shape of the surface is given by (\ref{rmu}) with $\beta$ equal to one and with
\begin{eqnarray}
  f_1 &=&   \frac{1424483}{1896036} m + \frac{8128}{132573} m^2 + \frac{14522}{96145} m^3 \nonumber \\
  k_1 &=&   \frac{9750}{97991} m^2 - \frac{139}{46715} m^3 \nonumber \\
  h_1 &=&   \frac{49427}{155163} m^3
\label{fkhConstsec}
\end{eqnarray}
These expressions for $f$, $k$, and $h$ at the surface are accurate to 10 significant digits.
For all practical purposes they are limited only by the uncertainty in the small
rotational parameter $m$, and of course by truncation of the
series at order $m^3$. Similarly, by (\ref{JvsFKH}), the surface conditions can be used to calculate the gravitational
coefficients in the external potential. The results for this special case are
\begin{eqnarray}
  J_2 &=&   \frac{156041}{931420} m - \frac{980}{25913} m^2 + \frac{339}{46459} m^3 \nonumber \\
  J_4 &=&   - \frac{6381}{56362} m^2 + \frac{2864}{63535} m^3 \nonumber \\
  J_6 &=&   \frac{5277}{46804} m^3
\label{fkhConst}
\end{eqnarray}
\section{Application to Planets}
In this section we apply the two-layer model of this paper
(with constant core and envelope densities) to the planets.
The model is a simple one for planets, and it is most applicable to
terrestrial planets and icy satellites that have a differentiated structure
consisting of either a metallic core and a rocky mantle or a rock, metal core and an icy mantle.
Application to the gas and ice giant planets will also be made, though a constant-density
envelope is not a very realistic model of these bodies.
However, with some generalization of the two-layer model to include envelopes
with arbitrary radial density profiles, application to giant planets can
be made much more realistic. The approximate theory of figures
approach presented in this paper is readily generalized to arbitrary
radial density profiles in the envelope, and the exact solution can also
be extended to this case.

\subsection{Earth}
Table~\ref{tab:earthmod} presents the eccentricities and gravitational coefficient $J_2$
for a two-layer model of the Earth with parameters $\rho_2/\rho_1
=0.401$, $Q_{\mathrm{V}} = 0.1674$, and $\epsilon_2 = 0.002$. Results are given
for the exact solution to the two-layer problem and for the theory
of figures.
Approximate solutions are valid to orders 1, 2, and 3 in the small
parameter $m$. 
Table~\ref{tab:earthmod} also lists the observed  values of $E_1$, $E_2$, and $J_2$.
Two-layer models provide a good match to the observed eccentricities
and an acceptable match to the gravitational coefficient. No attempt was
made to fine tune the model parameters.  For this case, even the
theory of figures to first order in $m$ gives good agreement with
the exact solution and with the observations.

\subsection{Mars}
Table~\ref{tab:marsmod}
 gives results for a Mars model with
$\rho_2/\rho_1 =  0.486$, $Q_V = 0.125$,
$\epsilon_2 = 0.00347$.  There are no observations of the eccentricity
of the Martian core-mantle boundary.  The models provide
good estimates of the eccentricity of the surface and the
theory of figures approximations match the exact solution of $E_2$
quite closely.  The eccentricity of the core-mantle boundary is
less than that of the surface, as was the case for the Earth models,
and $E_1$ from the theory of figures approximations agrees rather
well with the value of $E_1$ from the exact solution.
The model $J_2$ is not in particularly good agreement
with the observed $J_2$ for Mars, but it is emphasized that
we made no attempt to fine tune the model parameters to fit
$J_2$. Moreover, the radius of the Martian core and the densities
of the Martian core and mantle are not known.

\subsection{Neptune} 
Table~\ref{tab:neptunemod}
lists results for a Neptune model with
$\rho_2/\rho_1 = 0.157334$, $Q_{\mathrm{V}} = 0.091125$. $\epsilon_2 = 
0.0254179$ (parameter values based on a model in
C.Z.~\citet{Zhang:EMP1997}).
It is emphasized that the two-layer model with a constant-density
envelope is not a good model for an ice giant planet like Neptune.
Nevertheless, the shape of the surface is not too different
from Neptune's observed shape, but $J_2$ for the
model is almost a factor of 2 larger than the observed value.
We do not know if Neptune has a core-envelope configuration
or a continuous radial density profile.

\subsection{Uranus}

Table~\ref{tab:uranusmod} provides results for two Uranus models with
model parameters given in the table and based on
\citet{HoredHubba:EMP1983}.
As was the case for Neptune, the Uranus models do okay in matching
$E_2$ but fail to give good estimates for $J_2$.
Similar to Neptune, it is not known if the radial density
profile of Uranus is a smooth one or if it contains a
discontinuity associated with a core.

\section{Discussion and Conclusions}
The exact solution for the rotational distortion of a two-layer Maclaurin ellipsoid reported in
\citet{KongZhangSchub:JGR2010}
has been extended here to provide formulas for the standard spherical harmonic expansion of the external gravitational
field of the body. We have also presented a new approach to the evaluation of the theory of figures based on numerical
integration of ordinary differential equations.

The classical Radau-Darwin formula is a low order result from the theory of figures and its realm of validity
 has been evaluated for the two-layer model using the exact solution. It was found that the Radau-Darwin approximation is not valid for the rotational parameter
$\epsilon_2=  \Omega^2/(2\pi \mathrm{G}\rho_2)\geq 0.17$  since the formula predicts a surface eccentricity that is smaller than the eccentricity of the core-envelope boundary.
 Interface eccentricity must be smaller than surface eccentricity. For an envelope density of
 3000~kg~m$^{-3}$ the failure of the Radau-Darwin formula corresponds to a rotation period of about 3~hr.
 
The generalized Roche model, a two-layer model with an envelope density equal to zero, provides a simple model against which to evaluate the validity
 of the theory of figures against the exact solution. It was found that the theory of figures only slightly underestimates the eccentricities of the
 surface and core-envelope interface compared with the exact solution.
 
Application of the exact solution and the theory of figures is made to models of Earth, Mars, Uranus, and Neptune.
It is found that the two-layer model with constant densities in the layers can provide realistic approximations to terrestrial
planets and icy outer planet satellites. This is perhaps not surprising since the zeroth order structure of these planetary bodies
is similar to the two-layer model with constant densities in the layers. The situation is not as straightforward for giant planets
since a constant density envelope is not a particularly good representation of the density in the outer layers of such planets. 
However, the theory of figures, as developed in this paper, is readily generalized to models with arbitrary radial density profiles
in the envelope (though we have not carried this out in this paper). Such models will be particularly useful for Jupiter and Saturn
which might possess heavy element cores surrounded by gaseous envelopes. The envelope density can be represented by
polynomial functions of radius. Inversions of gravitational data based on these models provide constraints on the gas giant interiors independent
of assumptions about composition and equations of state. The exact solution for the two-layer Maclaurin ellipsoid can also be extended to
allow for a non-constant radial profile of envelope density. This is not as straightforward as the generalization of the theory of figures,
but it can be done. The solutions for two-layer bodies can therefore provide acceptable models for the rotational distortion of terrestrial,
gas giant, and ice giant planetary bodies.  These solutions can also serve as benchmarks to test the validity of complicated
numerical models that invert gravitational and shape data to infer the interior structure of planets.
\clearpage

\clearpage
\clearpage
\clearpage
\clearpage
\clearpage
\clearpage
\clearpage
\clearpage
\clearpage
\begin{center}
\setlength{\tabcolsep}{.4in}
\begin{table}
\caption{Eccentricities of the interface $E_1$ and the surface $E_2$
as a function of the rotation parameter $\epsilon_2$ for $Q_{\mathrm{V}} = 0.5$
and $\rho_1/\rho_2 = 2$. The surface eccentricity based on the
Radau-Darwin approximation is $E^{R-D}_2$. \label{tab:xy}}
\begin{tabular}{rrrr}
\multicolumn{1}{c}{$\epsilon_2$}& 
\multicolumn{1}{c}{$E_1$}& 
\multicolumn{1}{c}{$E_2$}&  
\multicolumn{1}{c}{$E_2^{R-D}$}\\
\hline 
0.01000& Ê 0.13959&  Ê0.14390& Ê 0.14383\\[-0.1in]Ê
0.02000& Ê 0.19761&  Ê0.20330& Ê 0.20288\\[-0.1in]Ê
0.03000&Ê  0.24158&  Ê0.24860&  Ê0.24782\\[-0.1in]Ê
0.04000& Ê 0.27886&   0.28670&  Ê0.28540\\[-0.1in]Ê
0.05000&  Ê0.31147&  Ê0.32010&   0.31824\\[-0.1in]Ê
0.06000&  Ê0.34050&  Ê0.35010& Ê 0.34769\\[-0.1in]Ê
0.07000&  Ê0.36770&  Ê0.37770&  Ê0.37453\\[-0.1in]Ê
0.08000&  Ê0.39250&Ê  0.40320& Ê 0.39931\\[-0.1in]Ê
0.09000&  Ê0.41592&Ê  0.42710&  Ê0.42238\\[-0.1in]Ê
0.10000&  Ê0.43790&  Ê0.44960&  Ê0.44402Ê\\[-0.1in]Ê
0.11000&  Ê0.45864&  Ê0.47090&  Ê0.46441\\[-0.1in]Ê
0.12000&  Ê0.47895&  Ê0.49130&  Ê0.48372\\[-0.1in]Ê
0.13000&  Ê0.49789&   0.51070&  Ê0.50208\\[-0.1in]Ê
0.14000&  Ê0.51604&   0.52930&  Ê0.51957\\[-0.1in]Ê
0.15000&  Ê0.53399&  Ê0.54730&  Ê0.53630\\[-0.1in]Ê
0.16000&  Ê0.55103&  Ê0.56460&  Ê0.55232\\[-0.1in]Ê
0.17000&  Ê0.56739&  Ê0.58130&  Ê0.56771\\[-0.1in]Ê
0.18000&  Ê0.58335&  Ê0.59750&  Ê0.58250\\[-0.1in]Ê
0.19000&  Ê0.59873&  Ê0.61320&  Ê0.59674\\[-0.1in]Ê
0.20000&  Ê0.61401&  Ê0.62854& Ê 0.61047\\[-0.1in]Ê
0.21000&  Ê0.62897&  Ê0.64350&  Ê0.62373\\[-0.1in]Ê
0.22000&  Ê0.64315&  Ê0.65800&  Ê0.63654\\[-0.1in]Ê
0.23000&  Ê0.65732&  Ê0.67224&  Ê0.64894\\[-0.1in]Ê
0.24000& Ê0.67128&   Ê0.68620&  Ê0.66093\\ [0.1in]
\hline
\end{tabular}
\end{table}
\end{center}%
\clearpage
\begin{center}	
\setlength{\tabcolsep}{.1in}
\begin{table}
\caption{Comparison of interface and surface eccentricities for several models with small envelope
densities computed from the exact theory and the theory of figures Roche model
evaluated to second order in $m$.  \label{tab:yz}}
\begin{tabular}{lclllcc}\\
\multicolumn{1}{c}{$\rho_2/\rho_1$}& 
\multicolumn{1}{c}{$\epsilon_2\rho_2/\rho_1$}& 
\multicolumn{1}{c}{$Q_v$}&  
\multicolumn{1}{c}{$E_1$}&
\multicolumn{1}{c}{$E_2$}&  
\multicolumn{1}{c}{$E_1^{\mathrm{ToF}} (\rho_2 = 0)$}&
\multicolumn{1}{c}{$E_2^{\mathrm{ToF}} (\rho_2 = 0)$}\\
\hline 
$10^{-2}$&     0.05&    0.5&       0.4297&    0.4629&    0.4268&   0.4603\\
$10^{-3}$&     0.05&    0.5&       0.4292&    0.46296&  0.4268&  0.4603\\
$10^{-4}$&     0.05&    0.5&       0.4291&    0.46296&  0.4268&  0.4603\\ [0.18in]
$10^{-2}$&     0.05&    0.33&     0.4378&    0.5189&    0.4268&  0.5141\\
$10^{-3}$&     0.05&    0.33&     0.4366&    0.5195&    0.4268&  0.5141\\
$10^{-4}$&     0.05&    0.33&     0.4365&    0.51955&  0.4268&  0.5141\\ [0.18in]
$10^{-2}$&     0.02&    0.5&       0.2730&    0.2953&    0.2724&   0.2949\\
$10^{-3}$&     0.02&    0.5&       0.27297&  0.29537&  0.2724& 0.2949\\
$10^{-4}$&     0.02&    0.5&       0.2728&    0.29537&  0.2724& 0.2949\\ [0.18in]
$10^{-2}$&     0.02&    0.33&     0.2757&    0.3314&    0.2724&  0.3313\\
$10^{-3}$&     0.02&    0.33&     0.2749&    0.3318&    0.2724&  0.3313\\
$10^{-4}$&     0.02&    0.33&     0.2748&    0.3318&    0.2724& 0.3313\\
\hline
\end{tabular}
\end{table}
\end{center}
\clearpage	
\begin{center}	
\setlength{\tabcolsep}{.2in}
\begin{table}
\caption{Application of exact solution and theory of figures to Earth.
\label{tab:earthmod}}
\begin{tabular}{lccccc}
\multicolumn{6}{c}{Two-layer Earth model: $\rho_2/\rho_1 = 0.401$,
$Q_{\mathrm{V}} = 0.1674$, $\epsilon_2 = 0.002$.}\\ \hline
& 
\multicolumn{1}{c}{1st order}& 
\multicolumn{1}{c}{2nd order}&  
\multicolumn{1}{c}{3rd order}&
\multicolumn{1}{c}{exact}&  
\multicolumn{1}{c}{observed}\\  \hline
$E_1$&             0.070765& 0.0707906& 0.0707907& 0.070593& 0.0707\\
$E_2$&             0.0810949&  0.0811186& 0.0811188& 0.081000& 0.082\\
$J_2(10^6)$& 1115.25&  1114.19&  1114.19& 1110.2& 1080\\
\hline
\end{tabular}
\end{table}
\end{center}
\clearpage
\begin{center}	
\setlength{\tabcolsep}{.2in}
\begin{table}
\caption{Application of exact solution and theory of figures to Mars.
\label{tab:marsmod}}
\begin{tabular}{lccccc}
\multicolumn{6}{c}{Two-layer Mars model: $\rho_2/\rho_1 = 0.486$,
$Q_{\mathrm{V}} = 0.125$, $\epsilon_2 = 0.00347$.}\\ \hline
& 
\multicolumn{1}{c}{1st order}& 
\multicolumn{1}{c}{2nd order}&  
\multicolumn{1}{c}{3rd order}&
\multicolumn{1}{c}{exact}&  
\multicolumn{1}{c}{observed}\\  \hline
$E_1$&                        0.0888247&     0.0888743&     0.0888747&   0.088859&    --\\
$E_2$&                        0.100246&       0.100294&        0.100295&     0.10030&   0.10837\\
$J_2(10^6)$&              1825.82&        1823.18&          1823.18&        1823.1&   1960.0\\
\hline
\end{tabular}
\end{table}
\end{center}
\clearpage
\begin{center}	
\setlength{\tabcolsep}{.2in}
\begin{table}
\caption{Application of exact solution and theory of figures to Neptune.
Observed values of eccentricity and gravitational coefficient are from
\protect{\citet{Linda:AstronJ1992}}
and 
\protect{\citet{Jacob:AstronJ2009}}.
\label{tab:neptunemod}}
\begin{tabular}{lccccc} 
\multicolumn{6}{c}{Two-layer Neptune model: $\rho_2/\rho_1 = 0.157334$,
$Q_V = 0.091125$, $\epsilon_2 =  0.0254179$.}\\ \hline
& 
\multicolumn{1}{c}{1st order}& 
\multicolumn{1}{c}{2nd order}&  
\multicolumn{1}{c}{3rd order}&
\multicolumn{1}{c}{exact}&  
\multicolumn{1}{c}{observed}\\  \hline
$E_1$&             0.143134&  0.143506&   0.143515& 0.15147&   --\\
$E_2$&             0.209326&  0.209642&   0.209658& 0.21019&  0.18414\\
$J_2(10^6)$&  6228.69&    6188.61&     6188.92&    6241.0&  3408\\
 \hline
\end{tabular}
\end{table}
\end{center}
\clearpage
\begin{center}	
\setlength{\tabcolsep}{.2in}
\begin{table}
\caption{Application of exact solution and theory of figures to Uranus.
Observed values of eccentricity and gravitational coefficient are from
\protect{\citet{Linda:AstronJ1992}}
and
\protect{\citet{Jacob:BAAS2007}}.
\label{tab:uranusmod}}
\begin{tabular}{lccccc}\\
\multicolumn{6}{c}{Two-layer Uranus model: $\rho_2/\rho_1 = 0.0883529$,
$Q_{\mathrm{V}} = 0.421875$, $\epsilon_2 = 0.103112$.}\\ \hline
& 
\multicolumn{1}{c}{1st order}& 
\multicolumn{1}{c}{2nd order}&  
\multicolumn{1}{c}{3rd order}&
\multicolumn{1}{c}{exact}&  
\multicolumn{1}{c}{observed}\\  \hline
$E_1$&            0.186917&  0.187284&   0.187296&   0.18752&  --\\
$E_2$&            0.207279&  0.207599&    0.207616&  0.20780&   0.212918\\
$J_2(10^6)$& 4847.54&    4812.63&      4812.62&     4821.4&     3341 \\  \hline 
&\\
\multicolumn{6}{c}{Two-layer Uranus model: $\rho_2/\rho_1 = 0.0791231$,
$Q_V = 0.0563272$, $\epsilon_2 = 0.0318902$.} \\  \hline
& 
\multicolumn{1}{c}{1st order}& 
\multicolumn{1}{c}{2nd order}&  
\multicolumn{1}{c}{3rd order}&
\multicolumn{1}{c}{exact}&  
\multicolumn{1}{c}{observed}\\  \hline
$E_1$&             0.115322&  0.115648&  0.115655&  0.14160&   --\\   
$E_2$&             0.213329&  0.213629&  0.213648&  0.21473&   0.212918 \\
$J_2(10^6)$& 5718.07&      5679.99&    5680.32&    5801.4&   3341 \\ 
 \hline
\end{tabular}
\end{table}
\end{center}
\clearpage
\appendix
\section{Level-Surface Coefficients for the Spheroidal Functions $f$, $k$ and $h$ to Order Three}
\label{AppendixA}
The radial coordinate $r$, normalized to the mean radius $s$, can be written as follows as a truncated power series to order three in $m$. 
\begin{eqnarray}
  \frac{r}{s} &=&   1 + m f \left( \frac{1}{3} - \mu^2 \right) + m^2 k \left( \frac{8}{15} - 4 \mu^2 + 4 \mu^4 \right) + \frac{1}{18} \left( m f \right)^2 \left( 4 -33 \mu^2 + 27 \mu^4 \right) \nonumber \\
&&   + m^3 h \left( \frac{26}{105} - 4 \mu^2 + 9 \mu^4 - 5 \mu^6 \right) + \left( m f \right)^3 \left( \frac{14}{81} - \frac{49}{18} \mu^2 + 5 \mu^4 - \frac{5}{2} \mu^6 \right) \nonumber \\
&&   + \left( m f \right) \left( m^2 k \right) \left( \frac{16}{63} - \frac{28}{15} \mu^2 + \frac{4}{3} \mu^4 \right) 
\label{rm}
\end{eqnarray}
This expansion for $r$ is equivalent to (\ref{rmu}), but with the powers of $m$ emphasized and stated explicitly.
With $k$ and $h$ set equal to zero, it is the expansion for an ellipse with flattening  $mf$, and with origin of coordinates at the center of the ellipse.

With this function for $r/s$, it is straightforward to derive the coefficients $C_{2j}^i$ for the potential functions $V _i$
to arbitrary order by means of (\ref{VSP}), and by means of the procedure used to derive the coefficients
$ C_{2j}^0$ in (\ref{C2j0}). For order one ($  i = 1$), the result for the spheroidal functions is obtained to order 2 in the form
\begin{eqnarray}
  C_0^2 &=&   \frac{2}{5} f  + \frac{13}{35} f^2 + \frac{8}{35} k \nonumber \\
  C_2^2 &=&   1 + \frac{4}{7} f + \frac{10}{7} f^2 - \frac{16}{35} k \nonumber \\ 
  C_4^2 &=&   \frac{36}{35} f + \frac{402}{385} f^2 - \frac{48}{385} k \nonumber \\
  C_6^2 &=&   \frac{12}{77} f^2 - \frac{96}{77} k
\label{C2j2}
\end{eqnarray}
and for order two, the coefficients to first order are
\begin{eqnarray}
  C_0^4 &=&   0 \nonumber \\
  C_2^4 &=&   \frac{20}{21} f \nonumber \\ 
  C_4^4 &=&   1 + \frac{200}{231} f  \nonumber \\
  C_6^4 &=&   \frac{50}{33} f
\label{C2j4}
\end{eqnarray}
For order three, there is only one non-zero coefficient to order one, $C_6^6$ which is equal to one.

The coefficients for the mass contribution exterior to the level surface at $\beta$ follow from (\ref{VSp}).
There is only one non-zero coefficient for order zero, the coefficient $C_0^{0 \prime}$ which is equal to one.
For order one the coefficients can be written as
\begin{eqnarray}
  C_0^{2 \prime} &=&   -\frac{4}{15} f - \frac{38}{315} f^2 - \frac{16}{105} k \nonumber \\
  C_2^{2 \prime} &=&   1 - \frac{8}{21} f + \frac{32}{105} k \nonumber \\ 
  C_4^{2 \prime} &=&   - \frac{24}{35} f - \frac{4}{55} f^2 + \frac{32}{385} k \nonumber \\
  C_6^{2 \prime} &=&   \frac{32}{77} f^2 + \frac{64}{77} k 
\label{C2j2sec}
\end{eqnarray}
and for order 2 as,
\begin{eqnarray}
  C_0^{4 \prime} &=&   0 \nonumber \\
  C_2^{4 \prime} &=&   -\frac{16}{21} f  \nonumber \\ 
  C_4^{4 \prime} &=&   1 - \frac{160}{231} f \nonumber \\
  C_6^{4 \prime} &=&   -\frac{40}{33} f
\label{C2j2}
\end{eqnarray}
and for order 3 there is just one non-zero coefficient $ C_6^{6 \prime}$ equal to one.

\section{Level-Surface Potential Functions}
\label{AppendixA1}
The coefficients $C_{2j}^i$ and $ C_{2j}^{i \prime}$ derived in Appendix~\ref{AppendixA}
can be substituted into (\ref{A2i}) for the potential functions. The internal normalized potential
$A_0$ on a level surface is obtained immediately to third order as
\begin{eqnarray}
  A_0 &=&   \left(1 + \frac{8}{45} f^2 + \frac{584}{2835} f^3 + \frac{64}{315} f k \right) S_0 + \left( \frac{2}{5} f + \frac{13}{35} f^2 + \frac{8}{35} k \right) S_2 \nonumber \\
&&   + S_0^\prime - \left( \frac{4}{15} f + \frac{38}{315} f^2 + \frac{16}{105} k \right) S_2^\prime + \left( \frac{1}{3} + \frac{4}{45} f + \frac{2}{189} f^2 + \frac{16}{315} k \right) m
\label{A0}
\end{eqnarray}
Similarly, the second degree potential function $A _2$ is obtained immediately by (\ref{A2i}),
but it is simplified somewhat by multiplying it through by 3/2. Because it must be independent of $\mu$ on a level surface,
and because it is multiplied by $P _2 \left( \mu \right)$, it is equal to zero. The final expression for $A_2$ is
\begin{eqnarray}
  A_2 &=&   \left( f + \frac{31}{42} f^2  + \frac{38}{63} f^3 - \frac{1}{7} h + \frac{4}{7} k + \frac{44}{105} f k \right) S_0 \nonumber \\
&&   \left( \frac{3}{2} + \frac{6}{7} f + \frac{15}{7} f^2 - \frac{24}{35} k \right) S_2 + \frac{10}{7} f S_4  +\left( \frac{3}{2} - \frac{4}{7} f + \frac{16}{35} k \right) S_2^\prime \nonumber \\
&&   -\frac{8}{7} f S_4^\prime - \left( \frac{1}{2} + \frac{10}{21} f + \frac{19}{63} f^2 + \frac{8}{15} k \right) m = 0
\label{A2}
\end{eqnarray}

Because $A_2$ is zero, a solution for the small rotational parameter $m$ can be found to third order. The result is
\begin{eqnarray}
  m &=&   \left( 2 f - \frac{3}{7} f^2 + \frac{20}{49} f^3 - \frac{2}{7} h + \frac{8}{7} k - \frac{584}{245} f k \right) S_0 \nonumber \\
&&   + \left( 3 - \frac{8}{7} f + \frac{524}{147} f^2 - \frac{32}{7} k \right) S_2 + \frac{20}{7} f S_4 \nonumber \\
&&   + \left( 3 - 4 f + 2 f^2 - \frac{16}{7} k \right) S_2^\prime - \frac{16}{7} f S_4^\prime
\label{em}
\end{eqnarray}
For the higher degree potentials, the coefficients $C_{2j}^i$ and $ C_{2j}^{i \prime}$ are substituted into (\ref{A2i}).
Then the above expression for m is substituted into the result, and terms higher than order three are dropped. In this way the
centrifugal potential enters explicitly only in $A_0$ and $A_2$. When this procedure is applied to $A_4$
and the result is multiplied through by 35/4, the final expression is,
\begin{eqnarray}
  A_4 &=&   \left( 3 f^2 + \frac{277}{77} f^3 - \frac{48}{11} h - 8 k + \frac{2152}{231} f k \right) S_0 \nonumber \\
&&   \left( 15 f + \frac{2385}{154} f^2 + \frac{156}{11} k \right) S_2 +  \left( \frac{35}{4} + \frac{250}{33} f  \right) S_4 
     + 16 k S_2^\prime +\left( \frac{35}{4} - \frac{200}{33} f \right) S_4^\prime = 0
\label{A4}
\end{eqnarray}
Similarly for $A_6$, with the result multiplied through by $-33/8$, the final result is,
\begin{eqnarray}
  A_6 &=&   \left( f^3 - \frac{10}{7} h + \frac{32}{7} f k \right) S_0 + \left( \frac{15}{14} f^2 + \frac{60}{7} k \right) S_2 \nonumber \\ 
&&   - \frac{25}{4} f S_4 - \frac{33}{8} S_6 + 5 f S_4^\prime - \frac{33}{8} S_6^\prime = 0
\label{A6}
\end{eqnarray}
Except for two obvious typographical errors in the $m$ term for $A_0$, these expressions for $A_0$, 
$A_2$, $A_4$ and $A_6$
agree with expressions given by
\citet{ZharkTrubi:PPI1978}.
They can of course be carried to higher order, either by introducing
higher-order spheroidal functions into (\ref{rmu}) or by extending 
(\ref{rLP}) to arbitrary order,
as carried out by Zharkov and Trubitsyn to fifth order 
\citep{ZharkTrubi:PPI1978}

\section{Evaluation of the Gravitational Moments}
The evaluation of the gravitational moments $S _{2 i}$ and $S _{2 i}^\prime$ that appear in the potentials $A _{2 i}$ is straightforward.
An expression for $r/s$ to arbitrary order is simply substituted into (\ref{phi}) and (\ref{phi2}) and the integration is carried out
 over $\mu$. The appropriate third-order expression for $r/s$ is given by(\ref{rm}), and the third-order expressions for the functions
 $ \phi_i$ and $ \phi_i^\prime$ evaluate to the following. They agree with expressions given by
 \citet{ZharkTrubi:PPI1978}.
\begin{eqnarray}
  \phi_0 &=&   1 \nonumber \\
  \phi_2 &=&   - \frac{2}{5} \left( f + \frac{1}{6} f^2 + \frac{2}{9} f^3 + \frac{4}{7} k - \frac{1}{7} h + \frac{4}{3} f k \right) \nonumber \\
  \phi_4 &=&   \frac{12}{35} \left( f^2+ \frac{1}{3} f^3 + \frac{8}{9} k + \frac{16}{33} h + \frac{40}{297} f k \right) \nonumber \\
  \phi_6 &=&   - \frac{8}{21} \left( f^3 + \frac{30}{143} h + \frac{192}{143} f k \right) \nonumber \\
  \phi_0^\prime &=&   \frac{3}{2} \left( 1 - \frac{4}{45} f^2 - \frac{244}{2835} f^3 - \frac{32}{315} f k \right) \nonumber \\
  \phi_2^\prime &=&   - \frac{2}{5} \left( f + \frac{9}{14} f^2 + \frac{8}{21} f^3 + \frac{4}{7} k - \frac{1}{7} h  + \frac{4}{7} f k \right) \nonumber \\
  \phi_4^\prime &=&   \frac{32}{105} \left( k + \frac{6}{11} h + \frac{14}{33} f k \right) \nonumber \\
  \phi_6^\prime &=&   - \frac{80}{1001} \left( h - 4 f k \right)
\label{phifcns}
\end{eqnarray} 

\section{Functions $G_{j i}$ for the Differential Equations}
As described in section~\ref{Sol}, the procedure for generating the ODE of (\ref{ODE}) sequentially produces
the right-hand side of the equations as functions $G _{j i}$ of $\beta$. The results of this process are
\begin{eqnarray}
  G_{2 1} &=&   0 \nonumber \\
  G_{4 2} &=&   3 \left( 1 - \frac{\delta}{S_0} \right) F_1^2 + \frac{1}{2} \beta \left( 2 - 9 \frac{\delta}{S_0} \right) F_1 F_1^\prime - \frac{1}{4} \beta^2 \left( 1 + 9 \frac{\delta}{S_0} \right) F_1^{\prime 2} \nonumber \\
  G_{2 2} &=&   \frac{4}{S_0} \left( 1 - \frac{\delta}{S_0} \right) \left ( F_1 + \beta F_1^\prime \right) - 5 \left( 1 - \frac{\delta}{S_0} \right) F_1^2 - 2 \beta \left( 2 - 3 \frac{\delta}{S_0} \right) F_1 F_1^\prime \nonumber \\
&&   -\frac{1}{3} \beta^2 \left( 4 - 9 \frac{\delta}{S_0} \right) F_1^{\prime 2} - 8 K_2 \nonumber \\
  G_{6 3} &=&   \frac{84}{5} \left( 1 - \frac{\delta}{S_0} \right) F_1^3 + \frac{14}{5} \beta \left( 2 - 9 \frac{\delta}{S_0} \right) F_1^2 F_1^\prime - \frac{7}{5} \beta^2 \left( 1 + 9 \frac{\delta}{S_0} \right) F_1 F_1^{\prime 2} \nonumber \\
&&   + \frac{24}{5} \beta \left( 4 - 3 \frac{\delta}{S_0} \right) K_2 F_1^\prime - \frac{8}{5} \beta \left( 2 + 9 \frac{\delta}{S_0} \right) F_1 K_2^\prime + \frac{8}{5} \beta^2 \left( 2 - 9 \frac{\delta}{S_0} \right) F_1^\prime K_2^\prime \nonumber \\
&&   - \frac{264}{5} \frac{\delta}{S_0} F_1 K_2 - \frac{3}{5} \beta^3 \frac{\delta}{S_0} F_1^{\prime 3} \nonumber \\
  G_{4 3} &=&   \frac{1}{2 S_0} \left( 1 - \frac{\delta}{S_0} \right) \left( 4 F_1^2 + 8 K_2 +6 \beta F_1 F_1^\prime + 3 \beta^2 F_1^{\prime 2} +8 \beta K_2^\prime \right) \nonumber \\
&&   -\frac{29}{5} \left( 1 - \frac{\delta}{S_0} \right) F_1^3 - \frac{1}{20} \beta \left( 62 - 159 \frac{\delta}{S_0} \right) F_1^2 F_1^\prime - \frac{2}{15} \beta^2 \left( 2 - 27 \frac{\delta}{S_0} \right) F_1 F_1^{\prime 2} \nonumber \\
&&   + 6 \left( 1 - \frac{\delta}{S_0} \right) F_1 F_2 +\frac{4}{5} \left( 5 + 31 \frac{\delta}{S_0} \right) F_1 K_2 + \frac{1}{2} \beta \left( 2 - 9 \frac{\delta}{S_0} \right) F_1^\prime F_2 \nonumber \\
&&   - \frac{2}{15} \beta \left( 134 - 63 \frac{\delta}{S_0} \right) F_1^\prime K_2 - \frac{2}{15} \beta^2 \left( 19 - 63 \frac{\delta}{S_0} \right) F_1^\prime K_2^\prime - \frac{1}{2} \beta^2 \left( 1 + 9 \frac{\delta}{S_0} \right) F_1^\prime F_2^\prime \nonumber \\
&&  \frac{1}{2} \beta \left( 2 - 9 \frac{\delta}{S_0} \right) F_2^\prime F_1 - \frac{2}{5} \beta \left( 2 - 21 \frac{\delta}{S_0} \right) K_2^\prime F_1 -12 H_3 + \frac{3}{5} \beta^3 \frac{\delta}{S_0} F_1^{\prime 3} \nonumber \\
  G_{2 3} &=&   \frac{8}{3 S_0^2} \left( 1 - \frac{\delta}{S_0} \right) \left ( F_1 + \beta F_1^\prime \right) \nonumber \\
&&   - \frac{2}{3 S_0} \left( 1 - \frac{\delta}{S_0} \right) \left( 5 F_1^2 - 6 F_2 + 6 \beta F_1 F_1^\prime + 3 \beta^2 F_1^{\prime 2} - 6 \beta F_2^\prime \right) \nonumber \\
&&   3 \left( 1 - \frac{\delta}{S_0} \right) F_1^3 + \frac{1}{5} \beta \left( 18 - 35 \frac{\delta}{S_0} \right) F_1^2 F_1^\prime - \frac{1}{15} \beta^2 \left( 22 + 45 \frac{\delta}{S_0} \right) F_1 F_1^{\prime 2} \nonumber \\
&&   - 10 \left( 1 - \frac{\delta}{S_0} \right) F_1 F_2 - \frac{8}{5} \left( 5 + 4 \frac{\delta}{S_0} \right) F_1 K_2 - 2 \beta \left( 2 - 3 \frac{\delta}{S_0} \right) F_1^\prime F_2 \nonumber \\
&&   + \frac{16}{15} \beta \left( 25 - 9 \frac{\delta}{S_0} \right) F_1^\prime K_2 + \frac{16}{15} \beta^2 \left( 1 - 9 \frac{\delta}{S_0} \right) F_1^\prime K_2^\prime - \frac{2}{3} \beta^2 \left( 4 - 9 \frac{\delta}{S_0} \right) F_1^\prime F_2^\prime \nonumber \\
&&  - 2 \beta \left( 2 - 3 \frac{\delta}{S_0} \right) F_2^\prime F_1 + \frac{16}{5} \beta \left( 2 - 3 \frac{\delta}{S_0} \right) K_2^\prime F_1\nonumber\\
&& + 12 H_3 - \frac{1}{45} \beta^3 \left( 16 + 33 \frac{\delta}{S_0} \right) F_1^{\prime 3} -8 K_3
\label{Gji}
\end{eqnarray}
These general equations, when applied to a particular problem, are not as complicated as they appear.
The application of Eq.~\ref{Gji}  to the two-zone model in Sec.~\ref{2zone} illustrates the method in more detail.
It illustrates our preferred method for application of the ODE approach to any interior calculation in general.

\end{document}